\documentclass[apj]{emulateapj}
\usepackage{apjfonts} 
\usepackage{lscape}

\usepackage{graphicx}

\newcommand{\masyr}{\hbox{$\; {\rm mas \ y}^{-1}\;$}}
\newcommand{\muasyr}{\hbox{$\; \mu{\rm as \ y}^{-1}\;$}}
\newcommand{\ba}{$\beta_\mathrm{app}\;$}
\newcommand{\bba}{\beta_\mathrm{app}\;}
\newcommand{\n}{\nodata}

\def\bi{\begin{itemize}}
\def\ei{\end{itemize}}
\def\be{\begin{equation}}
\def\ee{\end{equation}}

\def\gtrsim{\mathrel{\hbox{\rlap{\hbox{\lower4pt\hbox{$\sim$}}}\hbox{$>$}}}}
\def\lesssim{\mathrel{\hbox{\rlap{\hbox{\lower4pt\hbox{$\sim$}}}\hbox{$<$}}}}
\def\gtrsim{\mathrel{\hbox{\rlap{\hbox{\lower4pt\hbox{$\sim$}}}\hbox{$>$}}}}
\def\lesssim{\mathrel{\hbox{\rlap{\hbox{\lower4pt\hbox{$\sim$}}}\hbox{$<$}}}}

\received{July 28, 2009}
\accepted{September 28, 2009}
\journalinfo{Astronomical~journal}
\submitted{}

\shortauthors{Lister et al.}
\shorttitle{MOJAVE. VI. Kinematics of Blazar Jets}
\begin{document}
\title{MOJAVE: Monitoring of Jets in Active Galactic Nuclei with VLBA
Experiments.\\
VI.
Kinematics Analysis of a Complete Sample of Blazar Jets}
\author{
M. L. Lister\altaffilmark{1},
M. H. Cohen\altaffilmark{2},
D. C. Homan\altaffilmark{3},
M. Kadler\altaffilmark{4,5,6,7},
K. I. Kellermann\altaffilmark{8},
Y. Y. Kovalev\altaffilmark{9,10},
E. Ros\altaffilmark{11,9},
T.~Savolainen\altaffilmark{9},
J. A. Zensus\altaffilmark{9}
}
\altaffiltext{1}{
Department of Physics, Purdue University, 525 Northwestern
Avenue, West Lafayette, IN 47907;
\email{mlister@purdue.edu}
}
\altaffiltext{2}{
Department of Astronomy, California Institute of Technology, Mail Stop 249-17, Pasadena, CA 91125;
\email{mhc@astro.caltech.edu}
}
\altaffiltext{3}{
Department of Physics and Astronomy, Denison University,
Granville, OH 43023;
\email{homand@denison.edu}
}
\altaffiltext{4}{
Dr.\ Remeis-Sternwarte Bamberg, Universit\"at Erlangen-N\"urnberg,
Sternwartstrasse 7, 96049 Bamberg, Germany;
\email{matthias.kadler@sternwarte.uni-erlangen.de}
}
\altaffiltext{5}{
Erlangen Centre for Astroparticle Physics, Erwin-Rommel Str.~1,
91058 Erlangen, Germany
}
\altaffiltext{6}{
CRESST/NASA Goddard Space Flight Center, Greenbelt, MD 20771, USA
}
\altaffiltext{7}{
Universities Space Research Association, 10211
Wincopin Circle, Suite 500 Columbia, MD 21044, USA
}
\altaffiltext{8}{
National Radio Astronomy Observatory, 520 Edgemont Road,
Charlottesville, VA 22903-2475; 
\email{kkellerm@nrao.edu}
}
\altaffiltext{9}{
Max-Planck-Institut f\"ur Radioastronomie, Auf dem H\"ugel 69,
D-53121 Bonn, Germany;
\email{tsavolainen@mpifr-bonn.mpg.de, azensus@mpifr-bonn.mpg.de}
}
\altaffiltext{10}{
Astro Space Center of Lebedev Physical Institute,
Profsoyuznaya 84/32, 117997 Moscow, Russia;
\email{yyk@asc.rssi.ru}
}
\altaffiltext{11}{
Departament d'Astronomia i Astrof\'{\i}sica, Universitat de Val\`encia,
E-46100 Burjassot, Val\`encia, Spain;
\email{Eduardo.Ros@uv.es}
}

\begin{abstract}
We discuss the jet kinematics of a complete flux-density-limited sample
of 135 radio-loud active galactic nuclei (AGN) resulting from a 13 year
program to investigate the structure and evolution of parsec-scale jet
phenomena. Our analysis is based on new 2 cm Very Long Baseline Array
(VLBA) images obtained between 2002 and 2007, but includes our
previously published observations made at the same wavelength, and is
supplemented by VLBA archive data. In all, we have used 2424 images
spanning the years 1994--2007 to study and determine the motions of 526
separate jet features in 127 jets. The data quality and temporal
coverage (a median of 15 epochs per source) of this complete AGN jet
sample represents a significant advance over previous kinematics
surveys. In all but five AGNs, the jets appear one-sided, most likely
the result of differential Doppler boosting. In general the observed
motions are directed along the jet ridge line, outward from the
optically thick core feature. We directly observe changes in speed
and/or direction in one third of the well-sampled jet components in our
survey. While there is some spread in the apparent speeds of separate
features within an individual jet, the dispersion is about three times
smaller than the overall dispersion of speeds among all jets. This
supports the idea that there is a characteristic flow that describes
each jet, which we have characterized by the fastest observed component
speed. The observed maximum speed distribution is peaked at $\sim 10c$,
with a tail that extends out to $\sim 50 c$. This requires a
distribution of intrinsic Lorentz factors in the parent population that
range up to $\sim 50$. We also note the presence of some rare
low-pattern speeds or even stationary features in otherwise rapidly
flowing jets, that may be the result of standing re-collimation shocks,
and/or a complex geometry and highly favorable Doppler factor.
\end{abstract}

\keywords{
galaxies : active ---
galaxies : jets ---
radio continuum : galaxies ---
quasars : general ---
BL Lacertae objects : general ---
surveys
}
 \ 
\section{INTRODUCTION}

This is the sixth paper in a series in which we report results of the
multi-epoch Very Long Baseline Array (VLBA) MOJAVE program to
investigate the parsec-scale jet kinematics of a complete
flux-density-limited sample of 135 active galactic nuclei (AGN) in the
northern sky at 2 cm.  \cite{LH05} (Paper I) have described the
initial source sample selection and first VLBA polarization epoch
results, and in subsequent papers we have discussed the circular
polarization characteristics \citep{HL06} (Paper II), kpc-scale jet structure
\citep{CLK07} (Paper III), and parent luminosity function
\citep{CL08}(Paper IV) of the sample. In Paper~V
\citep{LAA09} we presented a large set of 2 cm VLBA images
obtained over a 13-year period beginning in 1994. A primary goal of
our study is a characterization of the long-term apparent jet motions
seen in these radio-loud AGNs.  In this paper, we discuss the
kinematic properties of the 135 AGN jets in the 
flux-density-limited MOJAVE sample (Paper~V) based on this
unprecedented large VLBA data set\footnote{The MOJAVE data archive is
maintained at http://www.physics.purdue.edu/MOJAVE}. We reported
preliminary findings regarding a correlation between jet speeds of the
MOJAVE AGN and their \textit{Fermi} $\gamma$-ray emission properties in
previous papers \citep{LHK09,KAA09}.  In subsequent papers we discuss
the polarization evolution, accelerations, and jet bending properties
of the sample.

The properties of a less well-defined sample of radio-loud AGNs
covering the period 1994 August to 2001 March and through 2006
September were discussed in earlier papers by \cite{KL04} and
\cite{CLH07}.  In these papers we reported apparent
jet velocities typically between zero and $15 c$, with a distribution
extending to about $35c$, and found that different features within
individual jets appear to move with comparable velocity.  In general,
we found that lower-luminosity AGNs have slower jets than the strong
blazars. We also noted the presence of apparently stationary jet
features, as well as the presence of bends and twists where the
observed motions do not back-trace to the base of the jet, but are
instead aligned with the local jet direction (i.e., they are
non-radial).  We found little or no evidence for backward motion
toward the jet origin, such as might be expected from random pattern
speeds.  However, detailed analysis was complicated by individual
features that may break up or combine with others in the jet, (e.g.,
\citealt*{K08}).

\cite{CLH07} have interpreted the observed speed distributions in terms of the
intrinsic bulk flow Lorentz factors and (unbeamed) synchrotron
luminosities, and reported results roughly consistent with simple
relativistic beaming models where the pattern velocity is comparable
to the flow velocity.  In a few cases we found clear changes in the
direction and/or speed of individual jet features as they propagate
along the jet (e.g., \citealt*{H03}).

We chose a wavelength of 2 cm for our VLBA observations to optimize
both resolution and sensitivity, and to minimize the impact of adverse
weather conditions.  Complementary radio-jet kinematic studies have
been reported at shorter wavelengths with better angular resolution
but lower sensitivity \citep{J01, J05} or at longer wavelengths with
better sensitivity to lower surface brightness features but poorer
resolution \citep{B07,B08, PMF07,KMO07}.

\cite{J01} used the VLBA at 7 mm and 1.3 cm to obtain multi-epoch
observations of 33 $\gamma$-ray blazars between 1993 and 1997.  Like
\cite{KL04}, they found that the jets of EGRET-detected $\gamma$-ray AGNs 
appear to be systematically faster than non $\gamma$-ray AGNs, and argued that
superluminal jet component ejection times occur at the time of
$\gamma$-ray outbursts.  Also using the VLBA, \cite{J05} reported
total and polarized 7 mm VLBA images of 15 AGNs at 19 epochs between
1998 and 2001 and derived Doppler and Lorentz factors using estimates
of intrinsic size estimated from the time scale of radio outbursts.

At 6 cm, \cite{B07} used an ad-hoc array and the VLBA to observe
293 AGNs from the Caltech-Jodrell Flat-Spectrum Survey
\citep{TVR96}, typically for 3 epochs spaced over the period 1990 to
2000. They reported on the distribution of velocities for 266 AGNs that
had good quality multi-epoch data and also found that the most luminous
sources tend to have the fastest apparent motions \citep{B08}.

\cite{PMF07} used data from the Radio Reference Frame Image
Database\footnote{http://rorf.usno.navy.mil/RRIFD} obtained with the
VLBA and up to seven other antennas at 3.4 cm to investigate jet
motions in 87 radio-bright AGNs during the period 1994--1998. Their
overall apparent speed distribution was peaked at low values, and
ranged up to $~\sim 30 c$.

The combined 2 cm VLBA Survey/MOJAVE data set presented here and in
Paper~V represents a significant improvement over prior studies in
terms of statistical completeness, temporal coverage (typically 15
epochs per source over 10 years) and data quality (typical image
resolution $\lesssim 1$ mas and rms image noise $\lesssim 0.4\;
\mathrm{Jy \; beam^{-1}}$).  In this paper, we exploit the longer time
range afforded by the newer observations as well as the use of
extensive archival VLBA data to study the distribution of jet speeds
in AGNs and to explore in more detail non-radial and accelerating jet
flows as well as the emerging evidence for low pattern speed features
found in otherwise rapidly flowing jets.  The overall layout is as
follows. In Section~\ref{dataset}, we discuss the observational set
and our method of defining the individual jet features. In
Section~\ref{kinematicssection}, we describe the jet kinematic
properties of the sample, and we discuss general statistical trends in
Section~\ref{overallstats}. We summarize our overall findings in
Section~\ref{summary}. We adopt a cosmology with $\Omega_m = 0.27$,
$\Omega_\Lambda = 0.73$ and $H_o = 71 \;
\mathrm{km\; s^{-1} \; Mpc^{-1}}$.  We use the convention $S_\nu \propto
\nu^{\alpha}$ for the spectral index. We refer to the radio sources throughout either by
their B1950 nomenclature, or commonly-used aliases.

\section{OBSERVATIONS AND DATA REDUCTION}
\label{dataset}
\subsection{Observational Data}

The data analyzed in this study were presented in Paper~V, and consist
of a set of 2424 2-cm VLBA observations of the complete
flux-density-limited MOJAVE sample of 135 AGNs spanning 1994 August 31
to 2007 September 6. Although the typical restoring beam of the
naturally-weighted images has dimensions of 1.1 by 0.6 mas, the
positions of individual jet features can be determined to much higher
accuracy, as we describe in Section~\ref{gaussianfitting}.  The
temporal coverage varies between 5 and 89 VLBA epochs per source, with
a median of 15 epochs. In scheduling our observations, attempts were
made to sample those AGNs with the most rapid jet motions more
frequently. Some sources had considerably more archival 2 cm VLBA data
(not from our program) owing to widespread community interest or their
popularity as calibrators.

The MOJAVE sample is selected on the basis of compact (2 cm VLBA) flux
density, which favors AGNs that have strongly beamed, fast jets viewed
at small angles to the line of sight (i.e., blazars).  The flux
density limit (1.5 Jy for declinations $\ge 0^\circ$, and 2 Jy for
$-20^\circ \le \delta < 0^\circ$) was chosen to provide a sufficiently
large number of quasars (N = 101), BL Lac objects (N = 22), and radio
galaxies (N= 8) to investigate statistical trends in the overall
sample and within the optical sub-classes.  The redshift information
on the sample is currently 93\% complete; the missing sources are
mainly BL Lacs with featureless spectra.  At the present time, only
four sources (0446+113, 0648$-$165, 1213$-$172, and 2021+317) lack
published optical counterparts.  The full list of MOJAVE sources and
their general properties can be found in Table~1 of Paper~V.

\subsection{Gaussian Model Fitting}
\label{gaussianfitting}

\begin{deluxetable*}{lclcrrcrc} 
\tablecolumns{9} 
\tabletypesize{\scriptsize} 
\tablewidth{0pt}  
\tablecaption{\label{gaussiantable}Fitted Jet Components}  
\tablehead{\colhead{} & \colhead {} &   \colhead {} & 
 \colhead{I} & \colhead{r} &\colhead{P.A.} & \colhead{Maj.} & 
\colhead{} &\colhead{Maj. P.A.}   \\  
\colhead{Source} & \colhead {I.D.} &  \colhead {Epoch} & 
\colhead{(Jy)} & \colhead{(mas)} &\colhead{(\arcdeg)} & \colhead{(mas)} & 
\colhead{Ratio} &\colhead{(\arcdeg)}   \\  
\colhead{(1)} & \colhead{(2)} & \colhead{(3)} & \colhead{(4)} &  
\colhead{(5)} & \colhead{(6)} & \colhead{(7)} & \colhead{(8)} & 
 \colhead{(9)}} 
\startdata 
0003$-$066  & 0& 2001 Jan 21  & 1.577  & 0.04 & 193.8 & 0.95 & 0.11 & 7\\ 
  & 1&   & 0.124  & 6.06 & 281.8 & 1.89 & 1.00 & \n\\ 
  & 2&   & 0.156  & 1.18 & 286.6 & 0.48 & 1.00 & \n\\ 
  & 7&   & 0.029  & 2.88 & 283.4 & \n & \n & \n\\ 
  & 0& 2001 Oct 31  & 1.280  & 0.02 & 217.5 & 0.87 & 0.11 & 9\\ 
  & 1&   & 0.146  & 6.49 & 284.0 & 1.32 & 1.00 & \n\\ 
  & 2&   & 0.125  & 1.27 & 292.6 & 0.49 & 1.00 & \n\\ 
  & 0& 2002 May 19  & 1.210  & 0.18 & 7.2 & 0.46 & 0.27 & 346\\ 
  & 1&   & 0.173  & 6.53 & 279.7 & 1.18 & 1.00 & \n\\ 
  & 2&   & 0.057  & 1.51 & 289.7 & 0.64 & 1.00 & \n\\ 
  & 3&   & 0.059  & 1.06 & 258.5 & 0.33 & 1.00 & \n\\ 
  & 4&   & 0.619  & 0.64 & 190.2 & \n & \n & \n\\ 
  & 7&   & 0.072  & 3.15 & 284.8 & 2.00 & 1.00 & \n\\ 
  & 0& 2003 Feb 5  & 2.349  & 0.05 & 204.7 & 0.89 & 0.25 & 2\\ 
  & 1&   & 0.204  & 6.65 & 281.0 & 1.25 & 1.00 & \n\\ 
\enddata 
\tablecomments{Columns are as follows: (1) IAU Name (B1950.0); (2) component name (zero indicates core component); (3) observation epoch; (4) flux density in Jy; (5) position offset from the core component (or map center for the core component entries); (6) position angle with respect to the core component (or map center for the core component entries);  (7) FWHM major axis of fitted Gaussian (milliacseconds); (8) axial ratio of fitted Gaussian; (9) major axis position angle of fitted Gaussian.   }
\tablenotetext{a}{Individual component epoch not used in kinematic fits.}
\end{deluxetable*} 

The majority of the jets in our sample have relatively simple radio
morphologies that consist of a bright (presumably stationary) ``core''
and a faint jet that typically extends for a few milliarcseconds on
one side of the core (Paper~V). Given the modest dynamic range of a
few thousand to one in our images, the jets can be typically
modelled as a series of distinct Gaussian features. For the purposes
of tracking these jet features across multiple epochs, we carried out
our analysis in the interferometric ({\it u,v}) plane by fitting the
brightest features in each source with two-dimensional Gaussian
components using the ``modelfit'' task in Difmap
\citep{1997ASPC..125...77S}. This method offers advantages over
fitting done in the image plane, as it uses the full resolution
capability of the VLBA
\citep{KKL05}. However, components defined in this way do not
necessarily correspond to real features in the image, but instead may
merely be mathematical constructs needed to reproduce a complex
brightness distribution.  In our previous analysis \citep{KL04}, we
examined each image for continuity in position, flux density, and
dimensions and fit Gaussian components in the image plane to determine
component positions, thus sacrificing effective resolution for robust
solutions. As discussed in Section~\ref{cptids}, in order to insure
against non-physical solutions in our kinematic analysis, as might be
introduced by ({\it u,v}) plane fitting, we have used only ``robust''
components with reliable identification across at least five
epochs. Our fits yielded positions, sizes and flux densities for
distinct features in the jets, which are listed in
Table~\ref{gaussiantable}.

An important constraint on our kinematic analysis is to be able to
reliably cross-identify individual jet features from one epoch to the
next, in order to trace their motions.  A major difficulty arises from
the fact that a model fit to an individual epoch is not necessarily
unique. Choices can be made for the number of components used, and
whether these are delta functions, circular Gaussians, or elliptical
Gaussians. We carried out the majority of our fits by initially
assuming an elliptical Gaussian for the core component, and circular
components for the jet components.  This was done to better fit
unresolved structure very near the cores, as is sometimes the case with
newly emerging components. We found that using circular Gaussians for
the jet components reduced the number of free parameters and gave more
consistent centroid positions across the epochs.

After each component was fit, we inspected the residual images for
regions along the jet above $\sim 5$ mJy and iteratively added new
components as necessary. If any component (including the core)
approached zero axial ratio or an extremely small size during the
iterations, we replaced it with a circular Gaussian initially
comparable in size to the restoring beam, and then if necessary, a
delta-function component. In order to maximize the level of continuity
across epochs, we used the model from the previous epoch (with
components restored back to initially circular or elliptical
Gaussians) as a starting point for the next epoch. Often it was
necessary to add or delete components as new jet features emerged and
others faded with time below our sensitivity level. We discuss the
errors in our fitted components in Section~\ref{errors}.

\subsubsection{Core Component Identification}

Radio-loud blazars typically have a pc-scale morphology that consists
of a faint jet and a bright, compact core component that is optically
thick at centimeter wavelengths. Because the phase self-calibration process removes
any absolute positional information from the data, the identification
of this component is crucial for kinematic studies, as it provides the
(assumed) stationary reference point for all the epochs. In the case
of the MOJAVE sample, most jets were highly core-dominated,
which greatly simplified the identification of the core component. It
was typically the brightest feature at the extreme end of a one-sided
jet at most or all epochs. However, in some of the two-sided jets
(discussed in Section~\ref{twosided}) and in a few individual sources
described below, the core component identification was not as
straightforward.

0202+149: This source contains  several jet components within $\sim
0.5$ mas of the bright eastern feature. High resolution 7 mm VLBA
images by \cite{MJM02} indicate a strong component 0.5 mas from the
core, which corresponds to the brightest feature in our 2 cm 
images. Since there is little evidence of a counterjet in VLBA images
published at other wavelengths, and the source is a core-dominated,
moderately high-redshift quasar, we assigned the core position to the
easternmost feature in the jet.

0212+735: This AGN ejected a very bright moving feature in early 1991
that has remained significantly brighter than the westernmost
feature. Previous 7 mm VLBA observations by \cite{L01a} confirm the
identification of the latter feature as the core.

0316+413 (3C~84): The central feature of this two-sided radio galaxy
contains complex, curved structure on scales smaller than 2 mas. There
is still uncertainty as to the exact location of the core within this
structure, even at shorter wavelengths \citep{DKR98}. We have
arbitrarily chosen the northernmost feature within the central complex
as the core component.

0333+321: There is significant small-scale structure near the
westernmost feature in the jet, both at 2 cm and at longer wavelengths
\citep{LS00}. Since there is no evidence of a counterjet
in this source at 2 cm or other wavelengths, we have assigned the
core identification to the westernmost feature, despite the fact that
is not the brightest feature at all epochs.

0738+313: Prior to 2001, the jet was dominated by a bright southern
feature, which subsequently faded. We consider the northernmost
feature to be the core of this source. 

0742+103: This is the only bona fide peaked-spectrum source in the
flux-density-limited MOJAVE sample \citep{TTT05}, and given the known
characteristics of this class, it is a potential candidate for
two-sided jet structure.  It is among the highest-redshift objects in
the MOJAVE sample ($z$ = 2.624). We have fit a slightly weaker component
to the southeast of the brightest feature at all epochs, but have
arbitrarily assigned the core position to the brightest feature in the
images. We were not able to find any published shorter wavelength VLBI
images that could be used to clarify the structure near
the core.

0923+392: This well-studied core-dominated blazar has a bright complex
of eastern jet components that have been interpreted as a region where
the jet bends directly into the line of sight. VLBI spectral analysis by
\cite{AKG97} has identified the (weaker) westernmost jet feature as the core.

1226+023 (3C 273): This well-known optically low-polarization quasar
frequently ejects very bright components that dominate the core
emission. We have assigned the core position to the northeastern-most
jet feature at all epochs. 

1228+126 (M87): This nearby radio galaxy has a relatively faint
diffuse extension to the east of the brightest feature in the jet,
which \cite{LWJ07} and \cite{KLH07} argued could be a counter-jet.  We have
assigned the core position to the brightest feature, since it has the
highest brightness temperature and is also the site of weak circularly
polarized emission \citep{HL06}. The kinematic fit that we present
here differs slightly from that of \cite{KLH07}, since we have not
used the 2005 April 21 epoch which lacked the St. Croix antenna.

1548+056: The north-south jet direction and poorer ({\it u,v})
coverage of this equatorial source make its core identification
difficult. We have assigned the core to the southernmost jet
feature.

1958$-$179: There are several components within the innermost 1 mas
of the jet. Given the difficulty in resolving the emission on this
small scale, we have assigned the core position to the brightest
feature at all epochs. 

2021+614: This radio galaxy has long been considered part of the
gigahertz-peaked spectrum class; however, owing to its long term
variability at high frequencies and shallow spectral curvature, we
have classified it as a flat-spectrum source. Archival multi-frequency
VLBA and space-VLBI observations \citep{LTM01} suggest the core is located in the
southernmost feature, which has a flat spectral index and high
brightness temperature. 

2128$-$123: This core-dominated, optically low-polarization quasar is
too strongly variable for the peaked spectrum class
\citep{TTT05}, despite having a radio spectrum that is slightly
peaked \citep{KNK99}. We have assigned the core position to the
northeastern-most feature in the jet.

2134+004: This source has a significantly curved ridgeline and highly
complex polarization structure \citep{LH05, HL06}. Shorter wavelength
VLBA observations by \cite{LS00} indicate that the jet core lies at the
southeastern edge of the radio structure.

2251+158 (3C 454.3): This powerful quasar undergoes frequent flares
and component ejections that dominate the core emission.  We have
assigned the core position to the easternmost jet feature at all
epochs.

\subsubsection{Jet Feature Identification Across Epochs}
\label{cptids}

In most cases, the identification of jet features across epochs proved
to be relatively straightforward, owing to the slow angular evolution of
the source with time. We also employed additional checks on the
continuity of flux density and brightness temperature over time to
confirm the cross-identifications. In some cases, however, the
situation was much more difficult because of one or more of the following
factors: a) insufficiently frequent temporal coverage, b) jet features
with extremely fast angular speeds and/or flux density decay
timescales, c) a high rate of new jet feature ejection, d) jets with
significant small scale (possibly curved) structure at or below the
VLBA resolution level, and e) jets containing both moving and
near-stationary features. \cite{PMF07} discuss how temporal
under-sampling can often lead to component misidentification and
incorrect estimates of apparent jet speeds in AGNs.

In Table~\ref{velocitytable}, we list all of the robust components that
we consider to have reliable identifications across a minimum of five
epochs.  We point out that despite the fact that some features in the
list of all fitted components (Table~\ref{gaussiantable}) have been
assigned the same ID number across epochs, only those listed in
Table~\ref{velocitytable} have robust cross-identifications. Many of
these were the result of comparing fits made completely independently
by two or more authors, who were free to make their own choices
regarding their fits and cross-identifications within the general
guidelines described above.

There were eight sources in the sample for which we were not able to
measure any robust jet speeds: 0109+224, 0235+164, 0727$-$115,
0742+103, 1124$-$186, 1324+224, 1739+522, and 1741$-$038. Three of
these jets were highly compact, with little resolved structure (see
Section~\ref{compact}). The sources 0235+164, 1324+224, and 1741$-$038
were all extremely compact, with no measured components at separations
larger than 0.4 mas. The components we did observe had considerable
scatter in their fitted positions, and none were robust.  Given that
the structure in these jets was on angular scales just at or less than
the size of the restoring beam, we cannot quote any reliable
upper limits on their apparent speeds. Their kinematics would be
better studied with shorter wavelength ground and/or space VLBI, which
could provide better angular resolution. The other five jets for which
we were unable to measure robust speeds were:

0109+224: This is a BL Lac object with an unknown redshift. A lower
limit of $z = 0.4$ was claimed by \cite{Falomo96} on the basis of its
optical host galaxy appearance.  The fitted jet components display an
unusual amount of positional jitter from epoch to epoch, which may
be a result of temporally under-sampled fast motions.

0727$-$115: There was too much positional scatter in the components to
consider any of them as robust. \cite{PMF07} analyzed 19 VLBI epochs at
4 cm and found three components with different speeds, all with
very large error values. \cite{KL04} reported a very high speed of $31.2
\pm 0.6 \;c$ but this was based on only three epochs at 2 cm.

0742+103: This source likely underwent a structural change in the core
region in the time period 2002-2004, during a temporal gap in our VLBA
observations. We are not able to measure any robust component speeds
in the jet because of uncertainties in the core position over time. The
parsec-scale jet structure is  approximately stable since 2005, which is
consistent with the low speed ($2.7 \pm 0.9 c$) reported by
\cite{KL04} at 2 cm. The three component speeds reported by \cite{PMF07} at 4 cm have
large error values.

1124$-$186: The jet structure in this low-declination source is very
compact and extended in a southern direction that unfortunately
coincides with the major axis of the restoring beam. The fitted
component positions thus suffer from high scatter, and we categorized
none of them as robust.  \cite{PMF07} reported two component speeds of
$1.8 \pm 29.5 \;c $ and $4.0 \pm 5.4 \;c$ for this jet at 4 cm.

1739+522: The jet is highly curved near the core, and contains
considerable small scale structure that made it difficult to
cross-identify its components across the epochs at the level of our
angular resolution. The single component reported by \cite{PMF07} at
4 cm had an inward speed of $16.8 \pm 7.9 \; c$.

\subsubsection{Errors on Fit Parameters}
\label{errors}

The errors in the model fit parameters of individual Gaussian
components in VLBA images are difficult to determine, because of 
non-linear dependencies on the interferometric array coverage, thermal
noise, and the presence of other nearby jet features.  For isolated
components, our previous work (see Appendix A of
\citealt*{HOW02}) has indicated typical errors of 5\% on the total
intensities and positional errors of $\sim 1/5$ of the restoring beam
dimension. For the brightest components, these errors are smaller by
approximately a factor of two.

The vector motion (i.e., constant angular velocity) and acceleration
models described in Section~\ref{accel} were fit to our position
versus time data using a $\chi^2$ minimization routine (Press et
al. 1992).  Given the well understood difficulties in obtaining robust
and statistically accurate component position uncertainties directly
from the interferometric data in each epoch independently, we followed
the approach used by \cite{HOW01} of initially assuming equal
weighting for all of our data points in a given fit, and then
uniformly rescaling their weights such that $\chi^2 =
N_\mathrm{points}-N_\mathrm{parameters}$.  In addition to the
limitation that all data points are equally weighted, this approach
has the disadvantage that the final, minimum $\chi^2$ value itself is
fixed to equal the number of degrees of freedom, and therefore it
cannot be used to determine the suitability  of the chosen motion
model.  However, under the assumption that the applied model is
appropriate, this approach does yield good uncertainty estimates in
all of the model parameters.

\begin{figure}[t]
\centering
\resizebox{1.0\hsize}{!}{
   \includegraphics[angle=-90]{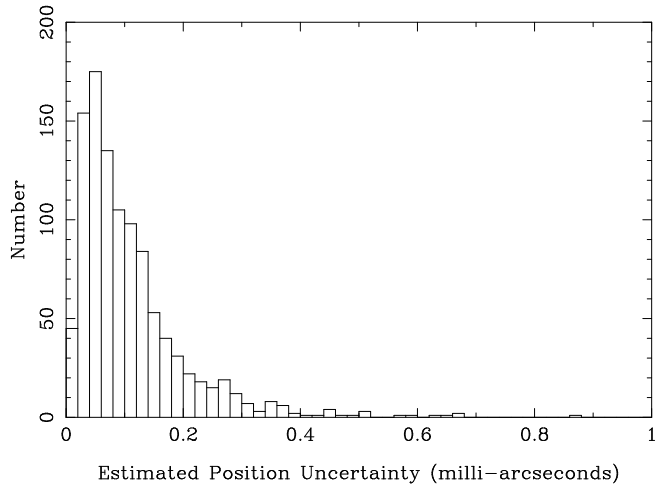}
}
\caption{\label{f:uncert_plot}
Histogram of the estimated typical position uncertainty 
for fitted Gaussian jet components. A total of 1052 component
uncertainties are plotted (2 for each of the 526 robust components in
our sample, one in the $x$-direction, the other in the $y$-direction).
The large $x$-position uncertainties for components 1 and 2 of
1228+126 (M87) ($1.01$ and $1.32$ mas respectively) are not plotted.
}
\end{figure}

We have studied the scatter of the measured component positions about
the best-fit proper motion model: this was either the initial vector
motion fit, or the acceleration model if the measured acceleration was
$\geq 3\sigma$.  Estimates of the typical component position
uncertainties appear in the last two columns of Table~\ref{velocitytable}.  In
Figure~\ref{f:uncert_plot}, we plot the distribution of these x and y
component position uncertainties.  The individual distributions for x
and y are so similar that we have simply combined the data to
construct Figure~\ref{f:uncert_plot}.  We find the most probable
component positional uncertainty is in the range $0.04-0.06$ mas, with 68\% of
the components having an estimated positional uncertainty $\leq 0.12$
mas and 95\% $\leq 0.29$ mas.

An alternate means of estimating the component positional errors
(e.g., \citealt*{PMF07}), is to examine the differences in the fits to
two or more individual epochs that are very closely spaced in time,
such that the jet has approximately the same structure in both data
sets. We have done this analysis on 10 pairs and one triplet of epochs
on several MOJAVE sources that were within seven days of each other,
and were of similar data quality. Based on the fastest observed proper
motion in each source, the expected real component movement between
the epochs in each of the pairs is less than 0.01\,mas for all the
sources except for BL\,Lac, which has an estimated movement of
0.04\,mas. Therefore, the differences in the fitted component
positions between the epochs should effectively reflect the positional
errors of the components.

The positional difference distribution of the individual fitted
components has a narrow Gaussian central peak at zero and non-Gaussian
tails up to roughly 0.6\,mas. In terms of their $x$ (R.A.) coordinates,
approximately 68\% of the components have a positional uncertainty
smaller than 0.08 mas, or roughly 10\% of the beam size.  In the $y$
(declination) direction, approximately 95\% of the components have a
positional uncertainty smaller than 0.33 mas (roughly 30\% of the beam
size).

We find that the positional accuracy depends on the
size of the component and, to some degree, on its flux density (i.e.,
signal-to-noise ratio). The positional difference distribution,
measured in units of component size convolved with the beam size, is
very close to Gaussian. For extended components, a 68\% (95\%)
confidence limit is 7\% (22\%) of the component size convolved with
the beam size. For a point source or delta-function component with
good SNR, the one-sigma uncertainty in position is less than one tenth
the beam size. These error estimates are consistent with those derived
from our multi-epoch fits as described above.

\section{PARSEC-SCALE JET KINEMATICS}
\label{kinematicssection}
\subsection{Vector Speeds}
\label{vectorspeeds}

For our initial velocity analysis of the robust features we have
assumed non-accelerating, two-dimensional motions on the sky. We fit vector
proper motions  to the $x$ and $y$ positions of all 526 robust
components in our sample, initially assuming simple, non-accelerating
motion.  For those features with at least ten epochs of data, we also
refit the data including acceleration terms, as described in Section
\ref{accel}, and in more detail by \cite{Homan09}. The resulting fits
are velocity vectors relative to the 
component's position at a reference (middle) epoch listed in column 14
of Table~\ref{velocitytable}. This table lists the magnitude, $\mu$,
of this vector (column 7) in microarcseconds per year, and its
direction on the sky, ($\phi$; column 9), as well as the least-square
fit errors on these quantities. Note that for components with $\geq
3\sigma$ accelerations, the values for $\mu$ and $\phi$ in
Table~\ref{velocitytable} are from the fit with the acceleration terms
included; however, as described is \S{3.2} these correspond closely to
the same values obtained from the unaccelerated fit.

In order to characterize whether the motion is directed purely
radially toward or away from the core, we compare $\phi$ with the mean
position angle of the component, ($\langle\vartheta\rangle$; column
6). We consider features in which the velocity is significantly offset
from the purely radial (i.e., $|\langle\vartheta\rangle - \phi| =
0$\arcdeg, or $|\langle\vartheta\rangle - \phi| = 180$\arcdeg )
direction at the 3 sigma level to be ``non-radial''. These are flagged
in column 10. We flag features as ``inward'' in column 9 if the
velocity vector offset is significantly in excess of $90\arcdeg$. We
discuss these rare inwardly moving features in more detail in
Section~\ref{inward}.

In column 13, we list the time of ejection (defined as when the core
separation equals zero) derived from the vector fit. No times of
ejection are listed for features in which any of the following apply:
i) the component shows significant acceleration (see Section~\ref{accel}),
ii) the magnitude of the vector motion is smaller than 3 times its
associated error, or iii) the motion is significantly non-radial or
inward.

\subsection{Accelerations}
\label{accel}

For the features in which we have obtained at least 10 epochs, there
is sufficient positional information to attempt a simple
two-dimensional acceleration fit.  The method is described by
\cite{HOW01}, and uses the following parameterization:

\begin{eqnarray}
x(t) &=& \mu_x(t-t_{x0}) + \frac{\dot{\mu}_x}{2}(t-t_{mid})^2 \\
y(t)& =& \mu_y(t-t_{y0}) + \frac{\dot{\mu}_y}{2}(t-t_{mid})^2 
\end{eqnarray}

\noindent Where $t_\mathrm{mid} = (t_{min}+t_{max})/2$ is the middle
epoch of our observations.  This epoch is chosen as the relative
epoch for fitting accelerations, and this choice has the benefit that 
the other parameters of the fit will correspond closely to 
those from the vector motion fit (which assumes no acceleration).  
Thus, the mean angular velocities in $x$ and $y$ are given 
directly as $\dot{x}(t_\mathrm{mid}) = \mu_x$ and 
$\dot{y}(t_\mathrm{mid}) = \mu_y$ respectively. The epochs of origin,
assuming these average speeds have applied since the component
emerged are also given directly by this fit: $t_{x0} 
= t_\mathrm{mid}-x(t_\mathrm{mid})/\dot{x}(t_\mathrm{mid})$ and $t_{y0} 
= t_\mathrm{mid}-y(t_\mathrm{mid})/\dot{y}(t_\mathrm{mid})$. 

\begin{figure*}[t]
\centering
\resizebox{1.0\hsize}{!}{
   \includegraphics[angle=270]{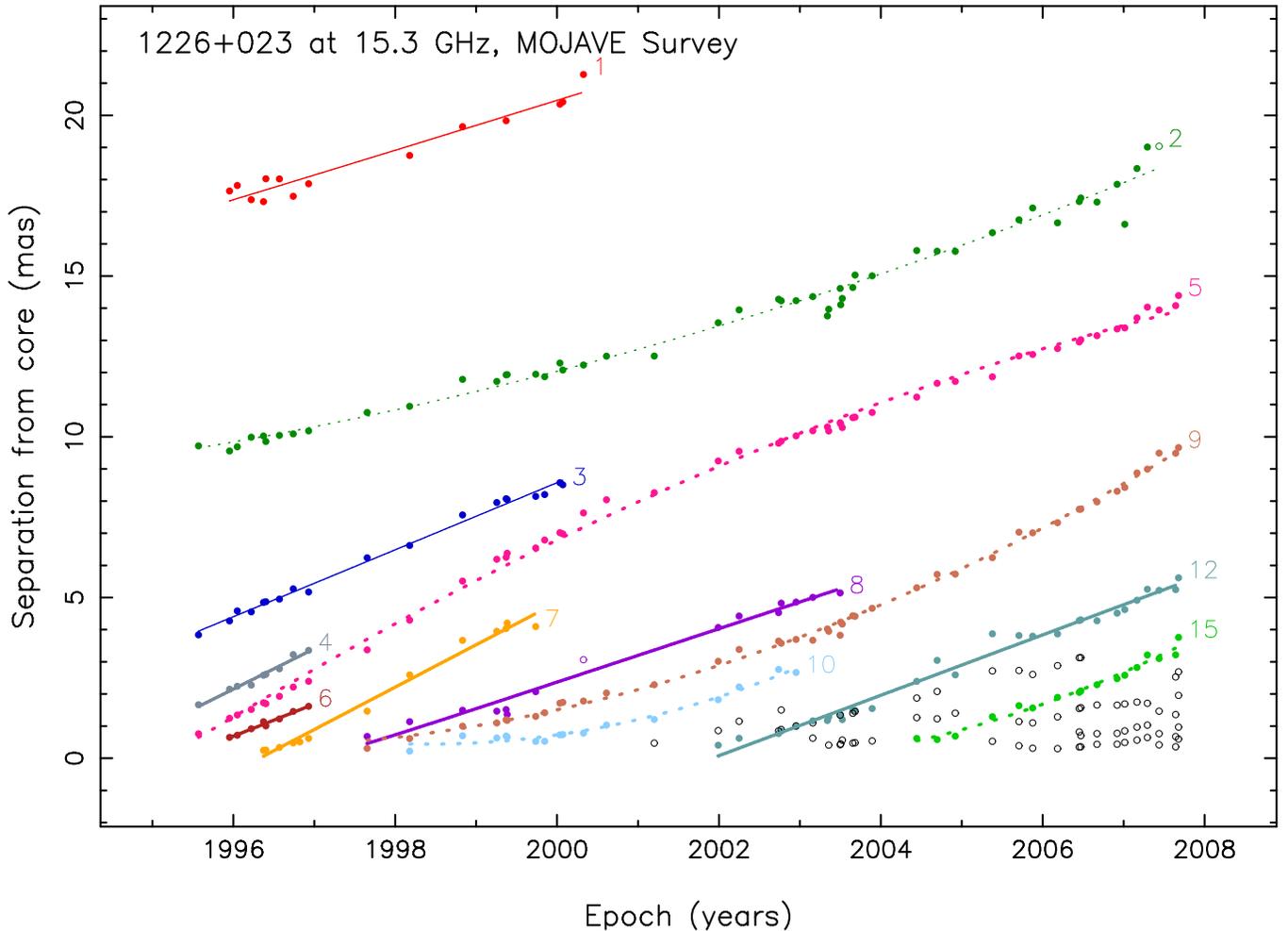}
}
\caption{\label{f:sepvstime}
Plot of angular separation from core versus epoch for Gaussian jet
components. The B1950 source name is given at the top left of each
panel. Color symbols indicate robust components for which kinematic
fits were obtained (dotted and solid lines). The solid lines indicate
vector motion fits to the data points assuming no acceleration, while
the dotted lines indicate accelerated motion fits. Thick lines are
used for components whose fitted motion is along a radial direction
from the core, while the thin lines indicate non-radial
motions. Unfilled colored circles indicate individual data points that
were not used in the kinematic fits, and unfilled black circles
indicate non-robust components.  The component identification number
is indicated next to the last epoch of each robust component. (This is
a figure stub; an extended version is available online.)  }
\end{figure*}

The angular accelerations on the sky plane are given by $\ddot{x} =
\dot{\mu}_x$ and $\ddot{y} = \dot{\mu}_y$. The accelerations are then
resolved into two components relative to the mean angular velocity
direction $\phi$: one parallel component, $\dot{\mu}_\parallel$, and
one perpendicular component, $\dot{\mu}_\perp$, along $\phi +
90^\circ$.  In this way $\dot{\mu}_\parallel$ directly quantifies
changes in apparent speed of the component during our observations,
and $\dot{\mu}_\perp$ quantifies changes in apparent direction on the
plane of the sky. In a few isolated cases where the angular speed of
the feature is extremely slow (less than $\sim 50\muasyr$, e.g.,
component 4 of 1253$-$055 (3C\,279), there may not be a well-defined
velocity vector direction. In such cases, the individual perpendicular
and parallel components of the acceleration are not necessarily
meaningful.

We note that our approach to fitting for angular accelerations assumes
that the same acceleration has applied throughout our observing
period.  Other models, such as impulsive accelerations inducing sudden
changes in speed and/or direction are certainly possible and may
provide a better fit in some cases, such as component 1 in 1253$-$055
(3C\,279) which was reported to make such a sudden change in
speed and direction by \cite{H03}.

We list the parallel and perpendicular components of the fitted
accelerations in columns 11 and 12, respectively, of
Table~\ref{velocitytable}. We have flagged in column 7 any features
which have a parallel or perpendicular acceleration exceeding the 3
sigma level. Of the 311 features we examined, 109 (35\%) fall into
this category. The angular speed ($\mu$) listed in
Table~\ref{velocitytable} for these features reflects the value
obtained from the acceleration fit, as opposed to the vector one.  It
is important to note that there are likely other jet features in
Table~\ref{velocitytable} that are in fact changing speed and/or
direction, but we cannot discern them as such because of insufficient
temporal coverage.

In Figure~\ref{f:sepvstime} we plot our fits to each source in a core separation
distance versus time diagram. Each robust jet component is indicated
by the colored points, which are joined by either a solid line
(non-accelerating fit) or a dotted line (accelerating fit).  Thin
line-widths are used in cases where the component motion is
significantly non-radial. Unfilled colored circles indicate individual
data points that were not used in the kinematic fits, and unfilled
black circles indicate non-robust components.

\begin{figure*}[p]
\centering
\resizebox{0.84\hsize}{!}{
   \includegraphics[angle=0]{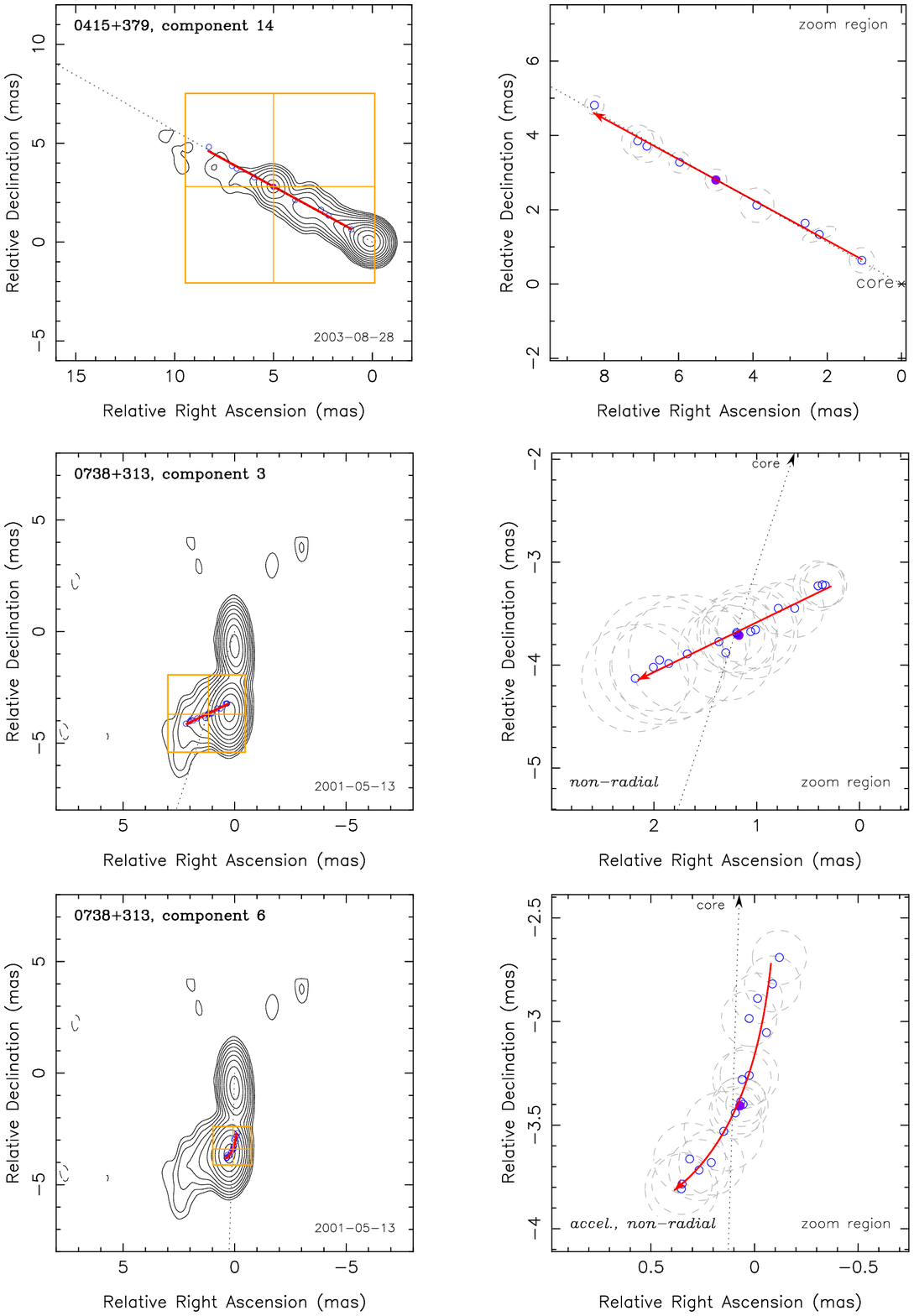}
}
\caption{\label{xyplots}
Vector motion fits and sky position plots of individual robust jet
features in MOJAVE AGN. Positions are relative to the core
position. The left hand panels show a 2~cm VLBA contour image of the
source from Paper~V at the middle epoch listed in
Table~\ref{velocitytable}. The orange box delimits the zoomed region
displayed in the right hand panels. The component's position at the
middle epoch is indicated by the orange cross-hairs. The dotted line
connects the component with the core feature and is plotted with the
mean position angle $\langle\vartheta\rangle$
(Table~\ref{velocitytable}). The position at the middle epoch is shown
by a filled violet circle while other epochs are plotted with unfilled
blue circles. The red solid line indicates the vector (or
accelerating) fit (see Table~\ref{velocitytable}) to the component
positions. The red arrows in the right hand panels indicate the
direction of motion, and the gray dashed circles/ellipses represent
the FWHM sizes of the individual fitted Gaussian components. Displayed
from top to bottom in the figure are component ID = 14 in 0415+379 (3C
111), ID = 3 in 0738+313, and ID=6 in 0738+313. }
\end{figure*}
\begin{figure*}[p]
\centering
\resizebox{0.84\hsize}{!}{
   \includegraphics[angle=0]{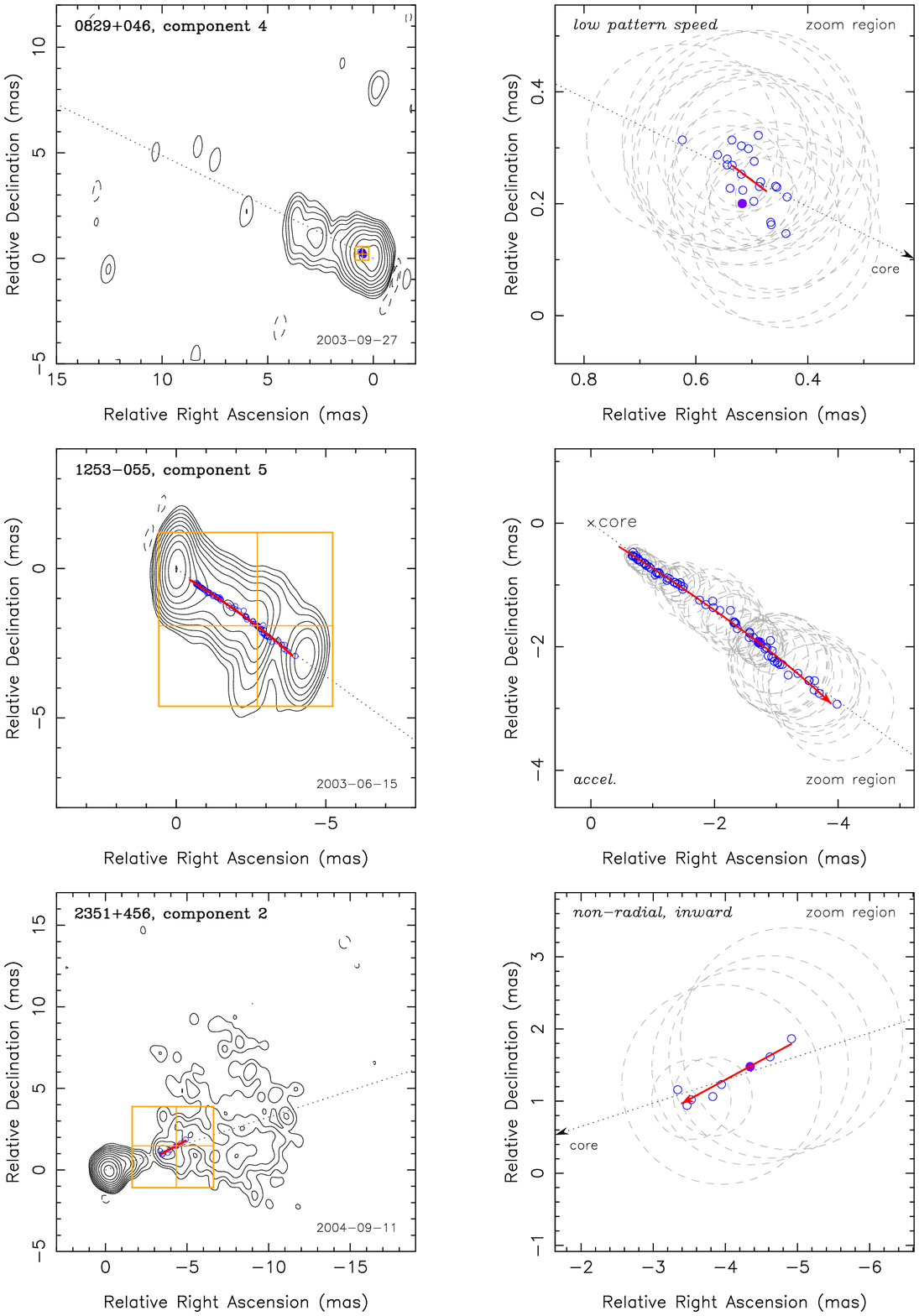}
}
\caption{\label{xyplots2}
Continued.  Displayed from top to bottom 
in the figure are component ID = 4 in 0829+046, ID=5 in 1253-055 (3C
279), and ID=2 in 2351+456. (This is a figure stub; an extended
version is available online.) 
}
\end{figure*}

We plot the motion fits and sky positions of the individual robust
components for selected AGN in Figures~\ref{xyplots} and
\ref{xyplots2}. Extended versions of Figures~\ref{f:sepvstime},~\ref{xyplots}, and
\ref{xyplots2}, containing plots of all of the robust jet 
components in the MOJAVE sample, are available online.  The left hand
panels of Figures~\ref{xyplots} and \ref{xyplots2} show a 2 cm VLBA
contour image  of the source from Paper~V at the middle epoch listed in
Table~\ref{velocitytable}. The orange box delimits the zoomed region
displayed in the right hand panels. The component's position at the
middle epoch is indicated by the orange cross-hairs. The dotted line
connects the component's middle epoch position with that of the core
feature. The positions at other epochs are plotted with unfilled blue
circles. The red solid line indicates the vector (or accelerating) fit
(see Table~\ref{velocitytable}) to the component positions. The red
arrows in the right hand panels indicate the direction of motion, and
the gray dashed circles/ellipses represent the FWHM sizes of the
individual fitted Gaussian components.

\subsection{Non-radial Motions}
\label{nonradial}

In our previous study of AGN jet kinematics in the 2 cm Survey
\citep{KL04}, we identified many cases where jet features were
moving in directions that did not extrapolate back to their presumed
origin at the base of the jet. These non-radial components provided
convincing evidence that jet features change direction, and ruled out
simple ballistic ``cannon ball'' models. In our study of the
MOJAVE sample, we find that 166 out of the 526 robust components (32
\%) satisfy our definition of non-radial motion
(Section~\ref{vectorspeeds}). A comparable fraction was found in
smaller samples by \citet{KL04} and
\citet{PMF07}, indicating that non-radial motion is a common feature
of the jet flow in blazars. We note that three of these non-radial
components might be considered ``inward'', i.e., their direction of
motion is greater than $90^\circ$ from the local jet direction at the
$3\sigma$ level.  Five additional components that would qualify as
inward do not qualify as non-radial since their
$|\langle\vartheta\rangle-\phi|$ values are less than $3\sigma$ from
180 degrees.  We discuss these features in more detail in
Section~\ref{inward}.

The high incidence of non-radial motion among the components in our
sample indicates that changes in jet motion are
common. A more detailed analysis of a subset of these data, to be
presented by \cite{Homan09}, indicates that in the
vast majority of cases, the non-radial motions are indeed caused by
accelerations in a direction perpendicular to the velocity vector.

\subsection{Apparent Inward Motions}
\label{inward}

For the great majority of our reported jet features, we find that
their motion is either directed outward from the core, or is nearly
stationary.  Fewer than ten of the 526 robust features we have studied
show evidence for inward motion, i.e., where the projected separation
between the jet feature and the core decreases with time,
and most of these are only marginally significant.  The illusion of
apparent inward motion can be explained in terms of one or more of the
following scenarios:

(i) If a newly emerging jet feature and the core are not resolved by
our beam, this would cause a temporary shift in the measured position
of the core centroid and a corresponding apparent decrease with time
in the apparent separation of the core and other more distant jet
features (e.g., \citealt*{GMM95}).

 (ii) It is also possible that the true core is not seen, possibly due
 to absorption, and that the only observed features are parts of a jet
 which are moving with different velocities (e.g.,
 \citealt*{Lobanov98}). If the furthest component is moving with a
 slower velocity than the one closest to the obscured core, then the
 two components will appear to be approaching each other.

 (iii) An apparent decrease in component separation from the core
 could also arise due to component motion away from the core but along
 a highly curved jet. If the jet bends back and across the line of
 sight, the projected separation from the core can appear to be
 decreasing, even though the actual distance from the core is
 increasing with time (e.g., \citealt*{MZS91}). 

(iv) Some observed jet features may merely reflect patterns in the
flow, of which some might even be moving backward. \cite{IP96} have
discussed an electron-positron jet model where an observer located
close to the direction of motion will see backward-moving knots.
\cite{G08} has proposed a model of a precessing jet which illuminates
an ambient screen, so that an observer oriented near the plane of the
screen can see projected motion in any direction even though the jet
itself is always flowing outward. We note, however, that any model
involving patterns must account for the the very large numbers of
observed outward motions, as opposed to inward ones.
 
(v) Internal changes in the brightness distribution of a large,
diffuse feature may cause the centroid of the fitted Gaussian to
wander, possibly mimicking inward motion.

We discuss several individual cases for inward motion in our sample
below:
 
1458+718: This compact steep-spectrum quasar has a very long jet with
a bright complex region located about 25 mas from the nucleus as well
as a fainter region about 40 mas downstream.  Close to the nucleus, as
well as for the more distant feature, the motion is in the outward
direction with \ba $\sim 7$.  It is more difficult to accurately
determine the motion of the more diffuse and complex feature located
about 25 mas downstream, but there may be a slight gross inward motion
with \ba $\sim 2$.
 
1803+784: This is a classical core-jet source.  The most distant, somewhat
extended, feature is moving away from the core with a
velocity of $9.0 \pm 2.5c$, and two intermediate features appear
almost stationary over the past ten years or more.  The two innermost
features each appeared to be moving away from the core with a velocity
of $\sim 1.5 c$ until the year 2000, when they both appeared to slow
down. Since 2004, the distance between of each of them and the core has
decreased.  The very similar apparent motion of these two inner
features is characteristic of an apparent change in separation due to
an emerging, yet unresolved new component. However, no such feature is
evident in VLBA data obtained as of March 2009.

2005+403: There is a pronounced bend in the jet located about 2 mas from the
nucleus.  Close to the nucleus, there is a bright feature which
has moved 0.1 mas closer to the nucleus over the three
year period that it has been observed, but this result depends
primarily on the first and last observations in 2004 and 2007
respectively.  Between mid 2005 and mid 2006 there was little or no
motion.  If real, the apparent inward motion of this feature is most
likely due to a new emerging component which is unresolved from the
core.  Beyond about 1 mas, the flow is in the general direction
of the jet ridge line.
 
2021+614: This jet has a slow-moving feature (ID = 2) located
approximately 3 mas downstream from the core that is classified as inward
according to our formal definition. The fitted angular speed is
extremely slow ($11 \pm 3$ \muasyr), and although there is a general
downward trend in its separation versus time plot, the component
oscillates about its mean position over time.

2200+420: The innermost component (ID=7) of this source changes
position angle from $-160$\arcdeg to $\sim -170 \arcdeg$ from 1999 to
2006, while maintaining a relatively stable separation of 0.3
mas. Since this is at the extreme limit of our resolution, and the
component has a low speed (Section~\ref{speeddispersion}), the vector
motion direction is not necessarily well-defined. We thus cannot
conclude for certain if the motion is truly inward.
 
2201+171: Close to the core, the flow is along the direction of the
jet, but 1.4 arcseconds downstream, a bright feature appears to flow in
the reverse direction, possibly because of confusion with a newly
emerging component. 

 2230+114 (CTA 102): This AGN has an elongated jet with several
 pronounced twists and bends.  Along most of the jet, the flow appears
 to be close to the jet direction, with apparent velocities between c
 and $16c$. A single component at $\sim 6$ mas from the core is moving radially inward at $0.97 \pm 0.22 c$. The contour maps
 indicate an apparently sharp bend of approximately $50^\circ$ in the
 jet at this location.

2351+456: As shown in Figure set~\ref{xyplots}, 2351+456 has a well
defined jet close to the core which has a sharp nearly right-angle
bend about 7 mas downstream, beyond which it opens into a very diffuse
structure characteristic of a direct head-on jet.  Close to the core
and near the bend the flow follows the ridge line with an apparent \ba
of 10 to 15.  However a bright feature located about 4 mas downstream
is moving toward the core with \ba = 28.3 $\pm$ 1.3.  Over the 6 year
period that we have observed this feature it has moved a total of
about 2 mas, so the apparent motion cannot be explained by the
emergence of a new feature unresolved from the core.  Rather, we
suggest that the jet of 2351+456 is viewed nearly end-on, and that the
region near 4 mas is where the jet bends back across the line of
sight. VLBA data from the MOJAVE program obtained after 2007 indicate
that the component at 4 mas has stopped moving inward, supporting this
interpretation.

\subsection{Dispersion of Apparent Speeds Within Individual Jets}
\label{speeddispersion}

Past multi-epoch VLBI studies of individual radio-loud AGNs have
identified instances where the jet displays features that move with a
range of apparent speeds (e.g., \citealt*{PMF07,J05,KL04}). Such
instances may suggest that the motions of some jet features are not
related to the flow, or that the jet is experiencing a change in
geometry over time due to precession.  Some jet features have also
been seen to either remain virtually stationary over time, or move at
significantly slower apparent speeds than other features in the same
jet \citep{KL04,Britzen05,J05}. These low-pattern speed (LPS) features
have sometimes been explained as discrete locations in a bent jet that
are at an extremely low viewing angle because of the specific geometry
involved \citep{AGM00}. In the specific case of a flow moving
relativistically along a helical path, the changing viewing angle is
predicted to give rise to Doppler boosted regions that appear nearly
stationary in VLBI images.  Alternatively, it is possible that some
LPS features represent standing shocks in the flow. These can take the
form of re-collimation shocks in an initially over-pressurized
outflow, and have been successfully reproduced in numerical
simulations of AGN jets (e.g., \citealt*{GMM95,PM07}). Some studies
(e.g., \citealt*{MC08,BDL08}) have suggested that standing shocks
within the unresolved cores of AGN jets may play a major role in
accelerating particles near the base of the jet, and could be
responsible for the persistent high levels of polarization in blazars
\citep{Darc07,MJ08}.

We have examined the distribution of apparent speeds within each jet in
the MOJAVE sample, and find a median rms dispersion of
$\mathrm{RMS}_\mathrm{jet}=0.065$~\masyr, or $2.6 c$. We also calculated
the median speed value of each MOJAVE jet, and find that the
distribution of this value for the entire sample has an rms dispersion
of $\mathrm{RMS}_\mathrm{sample}= 0.29$~\masyr, or $7.3 c$. The
$\mathrm{RMS}_\mathrm{sample}$ and $\mathrm{RMS}_\mathrm{jet}$ values
differ by more than a factor of three. This supports the previous
finding of \cite{KL04} that most jets generally eject features with a
characteristic speed.  We found even larger differences between
these two RMS values in the cases where i) the LPS features were dropped
from the analysis, or ii) only quasars were considered, or iii) the
maximum jet speeds were used to calculate
$\mathrm{RMS}_\mathrm{sample}$.

\begin{figure}
\centering
\resizebox{1.0\hsize}{!}{
   \includegraphics[angle=0]{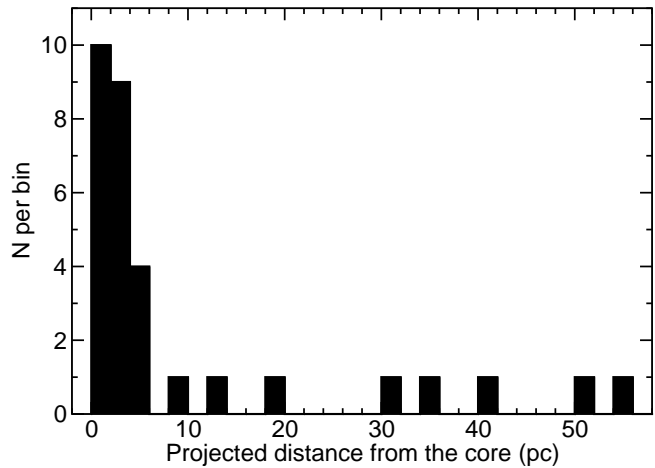}
}
\caption{\label{f:stationary_pcfromcore}
Distribution of projected distance from the core for low pattern speed
jet features in the MOJAVE sample.
}
\end{figure}

We have also used the regular time coverage and completeness of the
MOJAVE sample to investigate how prevalent LPS features are in
blazars, and whether they are more prone to occur in certain types of
AGN jets. We examined all 94 robust features in our sample with radial
speeds less than $50 \muasyr$ to see if their motions were
non-accelerating and significantly slower than other features in the
same jet. We were able to identify only 31 such instances, which are
indicated with a flag in column 5 of Table~\ref{velocitytable}. These
LPS features have a bimodal distribution, with the majority lying
within projected distances of $0.8$ mas ($\sim$6 parsecs) from the
core, while only a handful are found further downstream
(Figure~\ref{f:stationary_pcfromcore}).

Since the close-in LPS features are located approximately within one
beam-width of the core, we have investigated the possibility that they
may be artifacts arising from the  modelfitting process. We
model fitted the sources using different models for the core, including
circular and elliptical Gaussians, as well as delta-functions, and
were not able to eliminate the close-in features. They generally are
present in all the epochs on a particular source, except on occasions
when they are blended with new components that have emerged from the
core. After such components move downstream, the stationary feature
then reappears, indicating that it is a persistent  feature of the
flow.  We have also compared our model fits to five jets in common with
the \cite{J05} sample, who model fitted several closely-spaced VLBA
epochs at 7 mm with three times better angular resolution. In each
case, they identified a stationary component at roughly the same core
distance as in our fits (taking into account a typical core shift
between these wavelengths  of $\sim 0.1$ mas,
\citealt*{KLP08}). \cite{J05} found in some 
cases additional stationary features even closer to the core ($r
\simeq 0.1$ mas), which are below the resolution level of our 2 cm
images.  Despite our limited sensitivity to these extreme close-in
features, we conclude that LPS features are very rarely found in
blazar jets at projected distances greater than 6 pc downstream of the
core. Given the small viewing angles of the MOJAVE jets, this
translates into an approximate de-projected distance of 50 pc.

Given the caveats of our limited angular resolution, we find that the
incidence rate of LPS features is much higher for the BL Lacs in the
sample (10 of 22), as compared to the quasars (15 of 86). Although the
latter generally have higher redshifts than the BL Lacs, we find that
for the whole MOJAVE sample, K-S tests indicate no significant
differences in the redshift or 2 cm VLBA luminosity distributions of
the AGN jets with LPS features and those without. It is therefore
possible that the conditions in BL Lac jets are more conducive to the
formation of strong, stable re-collimation shocks. Numerical
hydrodynamical jet simulations by \cite{GMM95,GMM97} have shown that the
strength and separation of jet knots associated with re-collimation
shocks is dependent on the jet opening angle and Mach number, as well
as the external density gradient. 

\subsection{Two-sided Jets}
\label{twosided}

Although it is thought that quasars and radio galaxies have
intrinsically two-sided jets which feed the distant radio lobes, the
great majority of the sources in the MOJAVE sample have one-sided
jets, which is not unexpected considering the expected large
differential Doppler boosting of the approaching and receding jets.  A
few sources, however, show two-sided jets; these are mostly low
luminosity radio galaxies, and it is tempting to consider that these
sources lie close to the plane of the sky so that there is little
differential Doppler boosting.  However, \cite{CLH07} have shown that
the observed number of slowly moving low luminosity two-sided jets is
considerably greater than would be expected if the distribution were
determined by orientation alone, so that the two-sided jets may
reflect a population intrinsically different from the one-sided
population.

One challenge in determining independent speeds in a two-sided jet is
determining the unambiguous location of the common, presumably stationary,
central core presumed to lie at the base of each jet.  We have been able to
obtain robust speed measurements of both jet and counterjet components
for five two-sided jets in our sample.  Stacked images over all epochs
for these two-sided sources are shown in Paper~V and are discussed below.

0238$-$084 (NGC~1052): This is a bright nearby elliptical galaxy that
harbors a well-studied low luminosity AGN. \cite{VRK03,KL04,kad04b} have
previously discussed the observed outflow in two nearly symmetric jets
containing multiple features which propagate away from a central
region with relatively slow subluminal velocities.  VLA observations
show the jets extending symmetrically out to 1.5 kpc on either side
\citep{CLK07,Wro84}.  The central region is surrounded by a rich molecular
(OH and $\mathrm{H_2O}$), atomic, and ionized gas cloud.  HI observations
indicate the presence of three velocity streams, and multi-frequency
VLBA observations show patchy free-free absorption, which may come
from the torus surrounding the central black hole \citep{VRK03,k04b}.
The analysis of the jet motions is complicated by two factors. First,
the presumed base of the jets is obscured by free-free
absorption, especially on the receding side, and second, the
simultaneous presence of many features in each jet exacerbates the
problem of cross-identifying components over different epochs.  We
have estimated the location of a virtual center by assuming it is located
half way between pairs of symmetrically moving components in opposing
jets.  Figure set~\ref{xyplots} shows the location of each feature that we were
able to identify as a function of time.  Over the past decade, a new
feature is ejected at an average rate of about once
per year, each moving with an apparent velocity  in the range of
$0.1 c$ to $0.25 c$.

0316+413 (3C~84): NGC 1275 is a prominent radio galaxy that contains a
bright complex central core, a southern cocoon-like expanding cloud
which is connected to a central feature by a thin faint jet, as well
as an inverted spectrum counter feature \citep{DKR98,WDR00}. It was
detected in gamma-rays during the first 4 months of the \textit{Fermi}
observatory, during which time 2 cm MOJAVE VLBA observations indicated a
significant brightening of the core region \cite{Fermi3C84}.  We have
determined the motion in the southern lobe to be $\mu = 0.27\pm 0.05$
\masyr ($v = 0.31 \pm 0.06 \;c$) while the northern lobe
appears to move more slowly with $\mu = 0.07\pm 0.02 $ \masyr ($0.08
\pm 0.03 \; c$).  Assuming the jets are intrinsically symmetric, these
observed speeds correspond to a Lorentz factor of 0.6 with the motion
oriented at about 11 degrees to the line of sight. However, the
central component itself has a complex morphology, with observed
motions up to $0.2c$ or more \citep{DKR98}, which introduces some
additional uncertainty in the derived motions of the two outer
features.  We have taken the origin of motion to be at the northern
end of the central feature, as suggested by the 7 mm VLBA images and
corresponding motions within the central feature \citep{DKR98}.

1228+126 (Virgo A): M87 harbors a prominent jet which has been
imaged at radio, optical, and X-ray wavelengths. We have taken the
core to be coincident with the brightest radio component, which is
supported by the circular polarization reported at this position
\citep{HL06}. We are able to identify six features along the jet and
one counterjet feature which have existed continuously over the
decade-long interval covered by our observations.  As we reported
previously \citep{KLH07}, we find no evidence for any motion faster
than a few percent of light speed, with the fastest feature measured
at $0.05c \pm 0.02c$.  The counterjet may be moving at about $0.01c$,
but this is not well determined because of its complex structure. Faster
angular motions corresponding to $\sim 2 c$ have been reported by
\cite{Acciari09} in closely temporally-spaced 7 mm VLBA observations,
but these are still not sufficient to completely account for the large
jet-to-counterjet ratio in regions 3 mas downstream of the core by
Doppler boosting alone.  Therefore, the bulk plasma flow speed must be
much greater than the observed pattern motions \citep{KLH07}, and
there is the additional possibility that the M87 jet is intrinsically
asymmetric. We note that the extreme proximity and low luminosity
of M87 set it apart from the other AGN in our sample, which are
typically powerful FR-II class radio sources.

1413+135: This is a complex, highly unusual two-sided BL Lac object
located within an apparent spiral host galaxy \citep{VCV06}. Like
NGC~1052, the AGN is surrounded by a cloud of molecular,
atomic, and ionized gas \citep{PSC02}, but 1413+135 is a thousand
times more luminous than the AGN in NGC~1052.  We have identified
seven robust jet features in 1413+135, two of which are located toward
the north east in an apparent counterjet.  Both counterjet features
are weak, and so it is difficult to determine their motions with high
precision. Based on ten measurements over a seven year period, we find
a best-fit velocity of the more distant counterjet feature (ID = 1) of
$1.34 \pm 0.56c$, but we cannot exclude the possibility that this
feature is stationary.  The other counterjet feature (ID=2) is slow ($0.14
\pm 0.28$) and is also possibly stationary.  By comparison, in the
main jet, there are several bright, robust features with apparent
speeds ranging from near stationary for the strongest feature to $1.8
\pm 0.2c$ for the westernmost component (ID = 3). 

1957+405 (Cygnus A): Cygnus A is a classical (archetypical) FR II
radio galaxy.  The central compact radio source lies at the origin of
the well known prominent kiloparsec jet structure.  It has a
luminosity comparable to other powerful AGNs, but like the M87 AGN,
the Cygnus A AGN is many orders of magnitude less luminous than the
extended jets and lobes.  Also, as in M87, the jets are
asymmetric, with the western jet being brighter on scales of few
parsecs to a hundred kiloparsecs.  Interestingly, the ejection dates
of the counterjet components match up well with their counterparts in
the main jet, indicating near-simultaneous ejections.  On the bright,
presumably approaching side, the jet appears to be moving with a
velocity close to $0.2c$ and on the faint receding side,
$0.1c$. Assuming that the two sides are intrinsically similar, the
intrinsic velocity is about 20 percent of the speed of light, with the
jets oriented at an angle near 70 degrees to the line of
sight. Multi-wavelength VLBA studies by \cite{B05} report evidence for
an absorbing feature in front of the receding jet.

All of the apparent two-sided jets in our flux-density-limited sample
are found in relatively low luminosity and relatively nearby radio
galaxies.  We have not found any symmetric parsec scale jets in any
quasars, although 101 out of the 135 sources in our sample are
classified as quasars.  These two-sided jets provide additional
evidence that the apparent motions do indeed reflect the underlying
jet flow and are not simply a (shock) pattern motion which propagates
within a relatively stable or slowly moving plasma. In the case of
random pattern speeds, we should not expect motions in the jet and
counterjet to be correlated, whereas if the symmetric outflow is due
to a common explosive origin the kinematics of each side might show
similar properties. In at least NGC 1052 and Cygnus A we can make a
one-to-one association between features on the two sides, supporting
the idea that they reflect the real jet flow, although in M87 no such
association can be made. We note that in all of these sources, the
intrinsic flow speeds are only mildly relativistic, and the spatial
scales are much smaller than the more distant objects, thus we cannot
make any inferences from this argument about the nature of the
apparent superluminal speeds that are observed in the more powerful
quasars.

\subsection{Highly Compact Jets}
\label{compact}

We have also identified three extremely core-dominated sources (0235+164,
1324+224, 1741$-$038) which had very little discernible mas-scale jet
structure at any epoch. This could either be because their jets are
atypically weak, or they may possess a highly beamed core which swamps the
jet emission in our limited dynamic range images. It is also possible
that they may be experiencing a period of inactivity in which new jet
components are not being ejected. The latter was the case for
1308+326, which had very little discernible jet structure prior to
1997, and now has a pronounced mas-scale jet.

With the exception of 0235+164, none of the three sources show any
extended radio structure in deep VLA images
\citep{UJP81,PFJ82,CLK07}. They are therefore well-suited as
calibration sources for flux density and absolute polarization
position angle scaling when simultaneous single-dish/VLA and VLBA data
are available.

\subsection{Stable Jets}
\label{stable}

VLBI sources that maintain a stable mas-scale structure for long periods
with well-defined centroids are very useful in geodetic applications
such as the definition of the ICRF \citep{FV03}. Our long-term VLBA
study has indicated, however, that such sources are exceedingly rare. We
have identified only two sources in the MOJAVE sample whose components
have maximum apparent speeds smaller than 10 \muasyr. Both of these are
at high redshift. Unlike the compact sources discussed in
Section~\ref{compact} which may be in temporarily quiescent states, the
decade-long stability of these sources suggests that they have low
intrinsic expansion velocities and/or high jet viewing angles. However,
we cannot exclude the possibility that faster changes of the jet
structure of these quasars might occur in the future.

0552+398 (DA 193): This AGN (z = 2.363) has been frequently classified
as a gigahertz-peaked spectrum source in the literature. However, it
displays considerable flux density variability above its turnover
frequency \citep{TTT05}, and has persistent linear polarization on mas
scales that is uncharacteristic of the GPS class. We were able to fit
all of the 2 cm MOJAVE VLBA epochs between 1997 and 2007 with a two
component Gaussian model, in which the separation is barely increasing
at a rate of $4 \pm 1 $ \muasyr ($0.36 \pm 0.09 c$).

0642+449: This high redshift ($z=3.396$) quasar has also been previously
classified as a GPS, yet \cite{TTT05} found its flux density variability
to be too high for this class. During our 12 year monitoring interval we
have found a single robust jet component whose core separation is
increasing at only $7 \pm 1$ \muasyr ($0.76 \pm 0.11c$).

\begin{figure}
\centering
\resizebox{1.0\hsize}{!}{
   \includegraphics[angle=0]{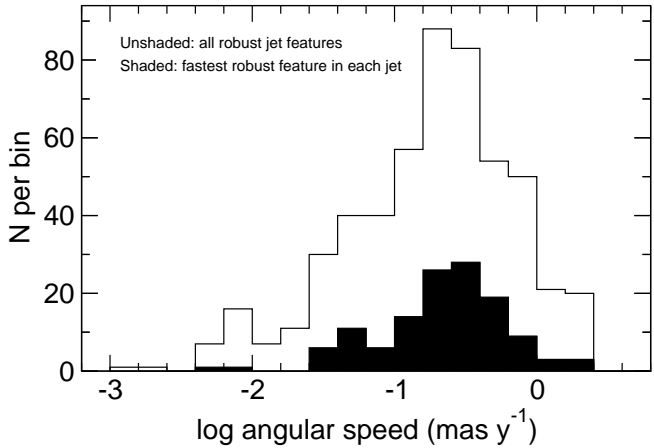}
}
\caption{\label{f:muasyrhist}
Angular speed distribution for all 526 robust
jet features in the MOJAVE AGN sample. The shaded histogram represents
the distribution of the fastest robust feature in each jet.  }
\end{figure}

\section{OVERALL KINEMATICS OF THE MOJAVE JET SAMPLE}
\label{overallstats}
\subsection{Distribution of Angular Jet Speeds}
\label{mudistrib}

The robust jet features in our sample span a wide range of apparent
speed, measured in both angular units and in units of the speed of
light. The distribution of angular speeds (Fig.~\ref{f:muasyrhist})
provides useful information for the planning of kinematic monitoring
programs (e.g., TANAMI: \citealt*{KMO07}), since it indicates how
often bright blazar jets should typically be observed to accurately
track their motions. The distribution is roughly log-normal, with a
median of 0.22\masyr.  The feature with the largest measured proper
motion is in the jet of the nearby radio galaxy 0430+052 (3C~120) and
is moving at $2.46 \pm 0.074$ \masyr. In the MOJAVE survey, we
have attempted to observe sources such that individual jet features move less
than a beam width between observations. This facilitates the
identification of features across successive epochs. Cadences range
from one observation per month for the fastest sources to once every
two to three years for the slowest sources, with a median of once
every 11 months (Paper~V).

\subsection{Maximum Jet Speeds}
\label{fastest}

A primary goal of the MOJAVE survey is to measure the distribution of
jet speeds in our flux-density-limited AGN sample, in order to
investigate the overall speed distribution and beaming properties of
the blazar parent population. Such information is needed for studies
of AGN luminosity functions and blazar demographics (e.g.,
\citealt*{CL08}). The issue is complicated somewhat by the possible
presence of multiple pattern speeds in an individual jet, some of
which may be moving at a different speed than the bulk flow
(Section~\ref{speeddispersion}). We are therefore interested in the
fastest speed in the jet measured over our approximately decade-long
monitoring period. For the purposes of constructing a maximum speed
statistic for each source, we considered all robust components in
Table~\ref{velocitytable} with angular speed measurements above the
$3\sigma$ level. This method addressed some isolated cases such as
2201+171, where a diffuse component with large positional
uncertainties gave a large (and poorly determined) speed measurement
that dominated the statistic for the source. In the case of 0422+004,
which had only a single robust component, we adopted its speed as the
maximum speed, despite its measured value being slightly below 3
$\sigma$.

\subsubsection{Distribution of Maximum Jet Speeds}
\label{speeddistrib}

\begin{figure}[t]
\centering
\resizebox{0.9\hsize}{!}{
   \includegraphics[angle=0]{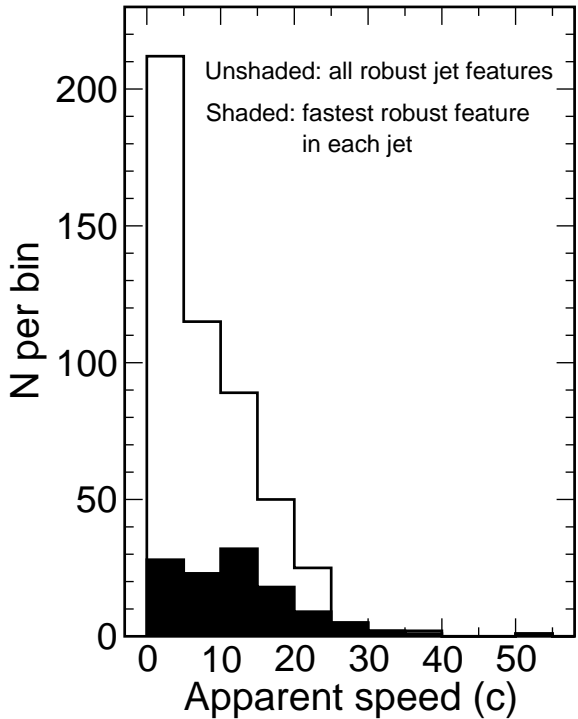}
}
\caption{\label{maxbetahist}
Distribution of apparent speed for 502 robust jet features in MOJAVE AGN
with measured redshifts. The shaded histogram represents the
distribution of the fastest robust feature in each jet.
}
\end{figure}

In Figure~\ref{maxbetahist}, we plot the distribution of jet speeds for
all robust features in the sample. The shaded histogram represents the
distribution of maximum speed. The fall-off towards high values
is consistent with previous large VLBI surveys carried out at longer
observing wavelengths \citep{B07, PMF07}. However, a major difference
in the MOJAVE distribution is that it is peaked at a moderately high
value (\ba $\simeq 10$), whereas the earlier surveys were peaked at
values $< 4 c$. The selection criteria of all three surveys are
similar, and contain many overlapping objects. The MOJAVE survey does
have higher angular resolution and excellent temporal coverage,
however, which may facilitate the identification of fast-moving
components that would otherwise be missed. Indeed, our fastest
identified jet component (component ID=3 in the quasar
0805$-$077\footnote{The redshift of 0805$-$077 ($z = 1.817$) was
originally measured by \cite{WJW88}, on the basis of several broad
emission lines in the optical spectrum, and was later confirmed at
the Nordic Optical Telescope by T. Pursimo (2009, private
communication).}) at $50.6 \pm 2.1 \,c$ is significantly faster than
any previously identified in cm-wavelength VLBA surveys, and is
comparable to some of the fastest components identified in a small
sample of 15 blazar jets observed monthly with the VLBA at 7 mm by
\cite{J05}. The median measured speed in the latter study was also
$\sim10 c$, which suggests that high angular resolution is essential
for determining accurate jet speeds in blazar proper motion surveys.
Data from the \textit{Fermi} observatory have shown that the extremely
fast MOJAVE blazars, including 0805$-$077 \citep{C09a,C09b}, show a
higher tendency to emit high energy $\gamma$-rays, likely as a result
of their high Doppler boosting factors \citep{LHK09,KAA09}.

\begin{figure*}[t]
\centering
\resizebox{1.0\hsize}{!}{
   \includegraphics[angle=-90]{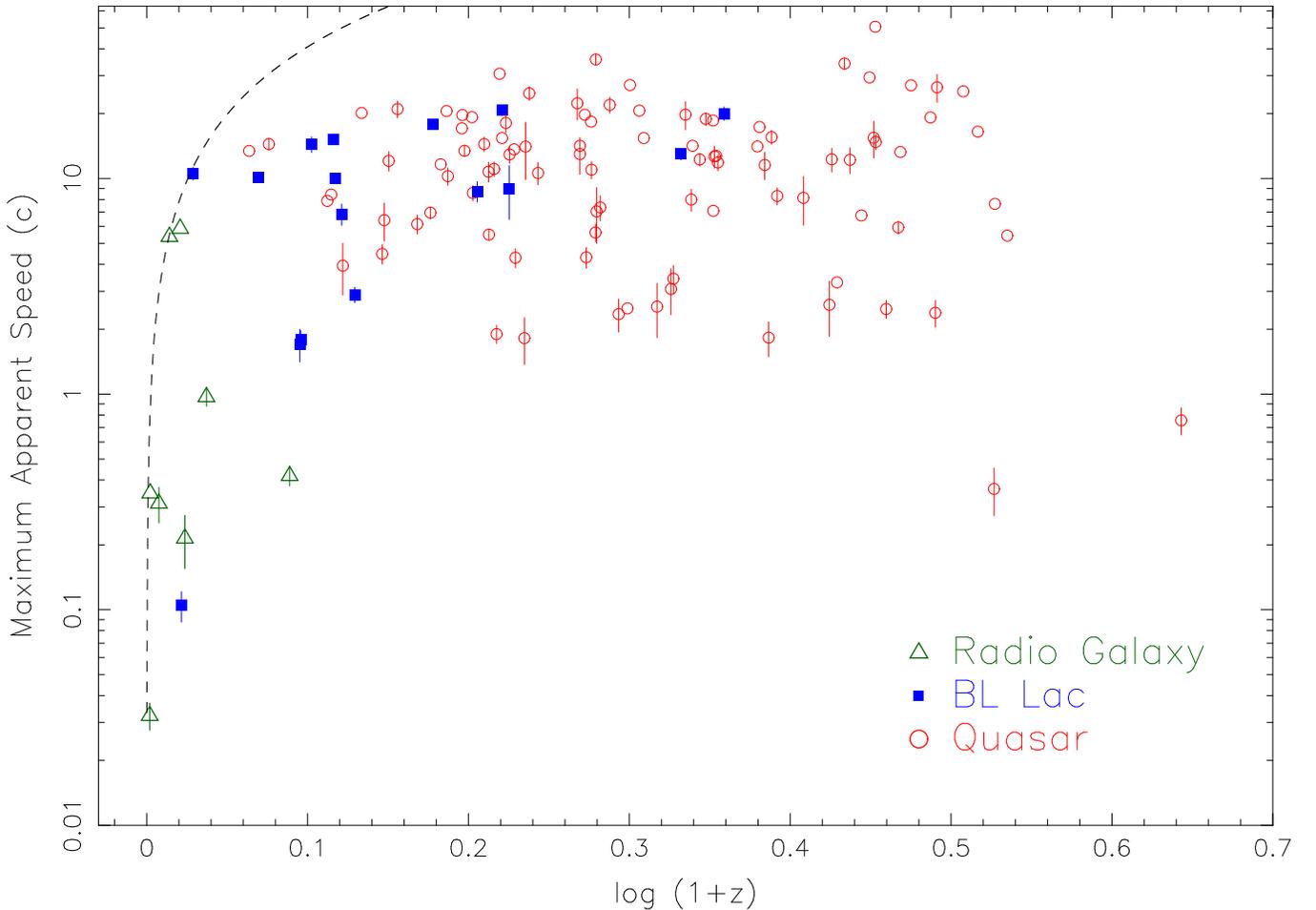}
}
\caption{\label{beta_appvsz} 
Plot of superluminal speed versus redshift for the fastest robust,
non-accelerating component in 119 MOJAVE jets. Error bars smaller than
the symbol sizes have been omitted. The red circles indicate quasars,
the filled blue squares BL Lac objects, and the green triangles radio
galaxies. The dashed curve represents an angular apparent speed of
2.5\masyr.
}
\end{figure*}

Other studies \citep{VC94,LM97} have investigated how the distribution
of fastest apparent speeds in an AGN sample selected on the basis of
beamed jet emission reflects the overall distribution of Lorentz
factors ($\Gamma$) in the parent population. First, the maximum
observed apparent speed sets an approximate value for
$\Gamma_\mathrm{max}$, since $\Gamma >
(\beta_\mathrm{app}^2+1)^{0.5}$, and in a large parent population,
some of the jets with $\Gamma =
\Gamma_\mathrm{max}$ will have their observed flux densities Doppler boosted
by very large factors by virtue of their near end-on orientations. For
reasonable choices of the parent luminosity function, most of these
sources will be selected in a flux-density-limited survey, regardless
of their redshift. Second, the shape of the distribution in
Figure~\ref{maxbetahist} rules out a single-valued $\Gamma$
distribution, since in that case the apparent speed distribution would
be expected to peak at \ba $\simeq \Gamma$, and drop off sharply for
smaller values of \ba \citep{VC94}.  The tapering of the speed
distribution at higher values strongly suggests a power-law Lorentz
factor distribution, which we use in Monte Carlo beaming simulations
described in Section~\ref{betalumsection}.

\subsubsection{Maximum Speed Versus Redshift}

In Figure~\ref{beta_appvsz}, we plot maximum speed against redshift for
the full sample. The data points are approximately evenly scattered for
redshifts greater than $\sim 0.6$, whereas at lower redshifts, there is
an apparent upper envelope, indicating that the maximum superluminal
speed increases with redshift. The dashed curve represents a constant
angular velocity of 2.5\masyr, which corresponds to the largest angular
speed measured in the survey. Since this curve lies well above the data
points past $z = 0.1$, we conclude that the upper envelope is not the
result of survey measurement bias.

The envelope instead likely arises because of the combined effects of
relativistic beaming and the sample's lower flux density cutoff. In the
MOJAVE survey, the minimum detectable luminosity rises sharply with
redshift, creating a classical Malmquist bias (see Fig.~1 of
\citealt*{CL08}). The high redshift sources therefore have higher
apparent luminosities, which they achieve primarily via Doppler
boosting. Our Monte Carlo simulations (Section~\ref{betalumsection})
suggest that these jets have some of the highest Doppler boosting
factors in the parent population (i.e., $\sim 2\Gamma_\mathrm{max}$),
and therefore can have apparent speeds up to $\bba =
\Gamma_\mathrm{max}$. At lower redshifts (below $z \simeq 0.6$), beaming
is not as important a factor in determining whether a jet makes it into
the sample, and the available co-moving volume is much smaller. This
makes it far less probable that highly superluminal jets will be seen at
low redshifts. The transition at which this occurs corresponds to a
redshift $z_o$ where the survey luminosity limit roughly coincides with
a flattening in the beamed luminosity function (LF). \cite{Lister03}
showed that the latter occurs at $L_1 \delta_\mathrm{max}^p$, where
$L_1$ is the low-luminosity cutoff of the intrinsic LF,
$\delta_\mathrm{max}$ is the maximum Doppler boosting factor, and $p$ is
the boosting index. For a flat spectrum jet with typical boosting index
of $p = 2$, $\delta_\mathrm{max} = 2 \beta_{\mathrm{app,max}} = 100$,
and $L_1 = 1.2 \times 10^{23} \; \mathrm{ W \, Hz^{-1}}$, which
corresponds to the faintest source in the survey (1228+126=M87), the
transition occurs at $z_o \simeq 0.55$, in approximate agreement with
Figure~\ref{beta_appvsz}.

\subsubsection{Maximum Speed Versus Apparent Radio Luminosity}
\label{betalumsection}

\begin{figure*}[p]
\centering
\resizebox{1.0\hsize}{!}{
   \includegraphics[angle=-90]{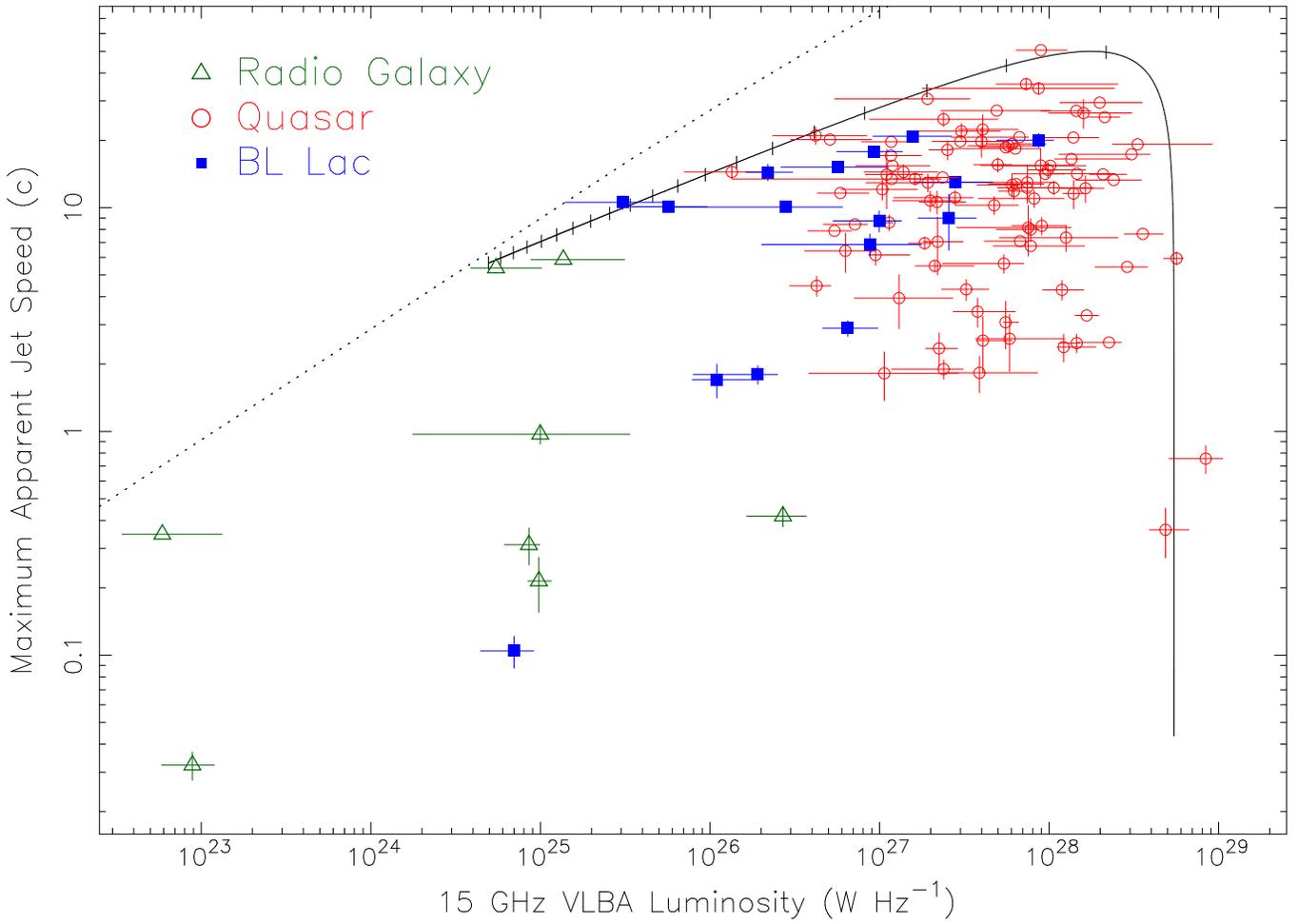}
}
\caption{\label{betalum}
Plot of maximum jet speed versus 2 cm VLBA luminosity for 119 jets in
the MOJAVE survey. The red circles indicate quasars, the filled blue
squares BL Lac objects, and the green triangles radio galaxies. Vertical
error bars smaller than the symbol sizes have been omitted. The
horizontal error bars indicate the range of luminosity of each source
during the period 1994.0--2004.0, and the symbols correspond to the
median values. The dotted curve corresponds to a jet with an apparent
speed of 2.5 \masyr and a flux density of 1.5 Jy, plotted over a range
of redshifts. The solid curve represents a jet at viewing angles ranging
from $5 \times 10^{-4}$ degrees to $20$ degrees, with Lorentz factor =
50 and an intrinsic unbeamed luminosity of $3 \times 10^{25} \; \mathrm{
W \, Hz^{-1}}$, Doppler boosted by a factor $\delta^{1.63}$. Ticks are
plotted along the solid curve at one degree intervals, starting at a
viewing angle of one degree and increasing towards the left.
}
\end{figure*}

In an earlier paper using preliminary MOJAVE data, \cite{CLH07}
reported an upper envelope in a plot of maximum jet speed versus VLBA
luminosity. In Figure~\ref{betalum}, we present a slightly different
version of this plot, where we have instead plotted the logarithm of
the jet speed in order to alleviate crowding close to the $x$-axis. In
calculating the luminosities, we have used the median total 2 cm VLBA
flux density measured during the period 1994.0--2004.0 and have
k-corrected the luminosities assuming a flat spectral index ($\alpha =
0$).

The dotted curve represents a jet with $\mu$ = 2.5 \masyr, flux density
of 1.5 Jy, and flat spectral index, plotted for a range of redshifts. 
These values correspond to the maximum angular speed measurement and the
flux density limit of the survey, respectively. The slope of this curve
is roughly $\sim 0.5$ in this log-log plot, since the apparent velocity
depends linearly on luminosity distance, whereas observed luminosity
varies as luminosity distance squared (the k-correction causes a slight
curvature at high $z$). As in the case of the maximum speed versus
redshift plot (Fig.~\ref{beta_appvsz}), the empty region between this
curve and the data points indicates that the envelope is not the result
of observational measurement limits.

\cite{CLH07} pointed out that the shape of the envelope was very similar
to the locus of points created by a single jet oriented over a range
of different viewing angles. The free parameters involved in fitting
this ``aspect curve'' to the envelope are the jet's unbeamed luminosity
(which determines its horizontal shift), its Lorentz factor (which
determines its shape and vertical extent), and the Doppler boost
index $p$, (which influences both the curve's shape and its slope at
low apparent luminosities).

For reasonably high values of the Lorentz factor, we
can derive the approximate slope of the left hand side of the
aspect curve, where $2/\Gamma \lesssim \delta << \Gamma$ (i.e., $\theta >> \sin^{-1}{(1/\Gamma)})$:
\begin{eqnarray} 
\label{bappeqn1} \beta^2_\mathrm{app} &=& 2\delta\Gamma - \delta^2-1 \\
\label{bappeqn2}	 & \simeq & 2\delta\Gamma. 
\end{eqnarray}
We also have
\begin{equation}
\label{eqnL} L = C_1 \delta^p,
\end{equation}
where $C_1$ is a constant.  Combining equations (\ref{bappeqn2}) and (\ref{eqnL}) yields 
\begin{equation}
\bba = \sqrt{2\Gamma}\; (L/C_1)^{(1/2p)},
\end{equation}
thus the slope in the log \ba vs. log L plot is $0.5\; p^{-1}$. 

The solid curve in Figure~\ref{betalum} represents a jet with Lorentz
factor $\Gamma = 50$ and an intrinsic unbeamed luminosity of $3 \times
10^{25} \; \mathrm{ W \, Hz^{-1}}$, viewed at angles ranging from $5
\times 10^{-4}$ to 20 degrees, and Doppler boosted by a factor
$\delta^{1.63}$. The tick marks are located at one degree intervals
moving from right to left, beginning at one degree from the line of
sight. The aspect curve provides a reasonably good fit to the
envelope, although there is considerable freedom in the fitting
parameters.  We determined the boosting index by performing a least
squares fit to the ten data points between $10^{24.7} \; \mathrm{W\; Hz^{-1}}
< L < 10^{28} \; \mathrm{W\; Hz^{-1}} $ that
lie along the upper envelope. This yielded a slope of 0.307 $\pm$
0.015, which corresponds to $p = 1.63 \pm 0.08$. This Doppler boost is
somewhat lower than that expected from a continuous flat-spectrum jet
($p = 2 + \alpha$, where $\alpha = 0$).

Although the aspect curve appears to provide a good fit to the upper
envelope, we find little justification for it in the framework of the
relativistic beaming model. One interpretation would require that
every jet located on the upper left edge of the envelope have a
Lorentz factor of 50 and an identical unbeamed luminosity. The lowest
apparent luminosity jets in this sub-group would then have viewing
angles of 10--20 degrees, and their Doppler factors would be on the
order of unity. However, since we are dealing with a
flux-density-limited beamed sample, there is a strong preference for
high-Doppler factor jets. At a given observed luminosity, e.g.,
$10^{26}
\; \mathrm{ W \, Hz^{-1}}$, there is a much higher probability of having
a jet with lower intrinsic luminosity and moderately high Doppler
factor, in part because of the steepness of the parent luminosity
function, and also because of the rarity of ultra-high Lorentz factor
jets in the parent population (see Section~\ref{speeddistrib}). Thus,
there is no a-priori reason to expect a sharp upper envelope bounded
by a curve with $\Gamma = \Gamma_{\mathrm{max}}$.  Another possible
explanation for the envelope, explored by \cite{LM97} and
\cite{CLH07}, is that there is a correlation between intrinsic jet
power and Lorentz factor, such that a jet with a particular $\Gamma$
must have a certain minimum unbeamed luminosity $L(\Gamma)$.

\begin{figure*}[p]
\center
\resizebox{1.0\hsize}{!}{
   \includegraphics[angle=0]{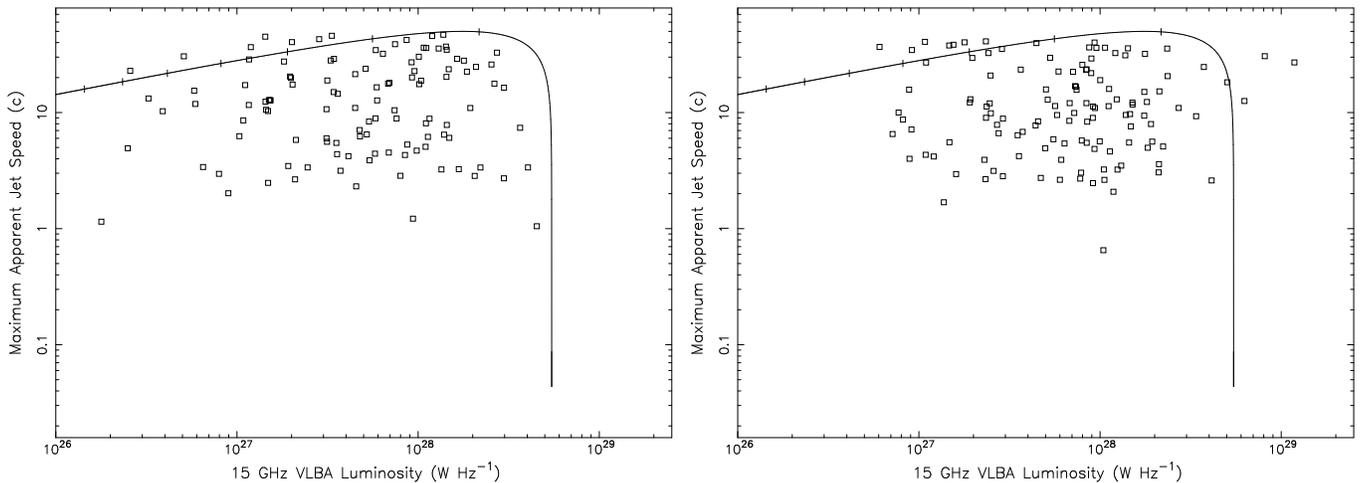}
}
\caption{\label{sim1} 
Left panel: plot of apparent speed versus observed luminosity for 135
simulated jets (see Section~\ref{betalumsection}). The solid curve is
identical to that in Figure~\ref{betalum}, however, the plotted $x$-axis
ranges of the figures are different. Right panel: identical simulation
to the left-hand panel, but with a different random number seed.
}
\end{figure*}

We have attempted to investigate these possibilities by performing
Monte Carlo simulations of a beamed parent population, from which we
select randomly generated jets above the MOJAVE flux density
threshold. The intrinsic jet luminosities are generated assuming a
power law pure-density evolution LF with slope -2.65 and redshifts
between 0.04 and 4, as described by \cite{CL08}. Because of the small
number of low-luminosity jets in the MOJAVE sample, these parameters
are chosen to optimize the fit to the quasar population. The jets are
assigned random orientations, and their Lorentz factors are drawn from
a power law distribution with slope = $-1.5$ and $ 3 <
\Gamma < 50$. The latter provides a reasonable fit to the overall
distribution of maximum jet speeds (Section~\ref{speeddistrib}). Each jet
is beamed using a Doppler boost index of $p = 1.63$. The simulation
is provided a random number seed, and jets are generated until a
flux-density-limited sample of comparable size to the MOJAVE survey is
obtained. With the choice of parameters above, and $10^{24} \;
\mathrm{W\; Hz^{-1}} <
L_\mathrm{intrinsic} <10^{27.6} \;\mathrm{W\; Hz^{-1}}$, this typically required
$\sim$ 20 million parent jets.

In Figure~\ref{sim1}, we show the results of two
simulations using these parameters, albeit with different random
number seeds. The fact that one of the simulations appears to
reproduce the envelope, while the other does not, reflects the
statistical fluctuations that are inherent in selecting a biased
subset of sources from a very large parent population. The apparent
flux density of any particular jet is determined by the joint
probability density of four variables, namely redshift, intrinsic
luminosity, viewing angle, and Lorentz factor. Thus, there are
multiple ways in which a jet can make it into a beamed sample, such as
having a low redshift, or a high Doppler factor, or both. Since the
flux density limit of the MOJAVE survey is shallow, we are likely
situated too far out in the statistical tail to fully constrain
particular aspects of the parent population, such as an intrinsic
relation between speed and luminosity. In order to address this issue,
we are obtaining speed information on an extended sample of AGN jets
that includes low-luminosity radio galaxies and new $\gamma$-ray
blazars detected by the {\it Fermi} observatory. 

\section{SUMMARY}
\label{summary}

We have investigated the pc-scale kinematic properties of 135 jets in
the complete  MOJAVE flux-density-limited survey of compact, radio-selected
AGN. The observational dataset consists of 2424 VLBA images at 2 cm
from our program and the VLBA archive between 1994 August 31 and 2007
September 6, and represents a significant improvement over previous
multi-epoch AGN surveys in terms of statistical completeness, temporal
coverage, and overall data quality.

Our main results can be summarized as follows:

(i) By modelling individual bright jet features with Gaussian
components, we were able to robustly identify and track a total of 526
moving features in 127 AGN jets. Three AGN (0235+164,
1324+224, 1741$-$038) had very little discernible mas-scale jet
structure at any epoch, while five others displayed too much scatter
to reliably cross-identify their features across multiple epochs. We
have also identified two bright AGN (0552+398 and 0642+449) with
extremely stable milliarcsecond-scale structures that have changed by
less than 10 \muasyr over a 12 year interval.

(ii) We fit a simple linear model to the positions of each component
over time in order to determine its speed and direction with respect
to the (presumed stationary) core feature located near the base of the
jet. Approximately 1/3 of the components had velocity vectors which
did not extrapolate back to the core feature. A more detailed analysis
presented elsewhere \citep{Homan09} links these 'non-radial' motions to
accelerations perpendicular to the velocity vector.

(iii) The jet features show an overwhelming tendency to display
outward motions, away from the core (98.5\% of all components). Our
analysis revealed only 8 inward-moving components, and in only one
case (ID=2 in 2351+456) is the motion highly superluminal. VLBA data
obtained since 2007 on the latter feature has indicated that its
inward motion has stopped, suggesting that it may be a case where the
jet has crossed our line of sight.

(iv) Features within an individual jet tend to have a modestly narrow
distribution of apparent speed (three times smaller than the overall
speed dispersion in the sample), supporting the idea that they are
connected with the intrinsic flow. We identified only 31 features
whose motions were slow and significantly smaller than others in the
same jet. These ``low-pattern speed'' features are more prevalent in
the BL Lac objects (10 of 22) than in the quasars (15 of 86), and are
rarely found at de-projected distances greater than 50 pc downstream
of the core. They may represent strong, stable re-collimation shocks
of the type commonly seen in numerical jet simulations.

(v) We have analyzed the kinematics of five MOJAVE sources that have
jets of similar brightness located on either side of the central
AGN. We find that at least two (NGC 1052 and Cygnus A) of these
two-sided jets display a close similarity in the ejection epochs and
apparent speeds of their jet and counterjet, providing additional
evidence that the moving features are related to the flow. 

(vi) Nearly 60\% of the 526 robust components in the sample had at
least 10 epochs of measurement; for these we also performed a
two-dimensional fit to determine their velocities and
accelerations. Approximately one third of these display significant
accelerations at the $3\sigma$ level, indicating that changes in speed
and/or direction are a common feature of AGN jets.

(vii) We have compiled a maximum speed statistic for each jet, which
is equal to the fastest robust feature observed in the source. The
distribution of these maximum speeds is peaked at $\bba \simeq 10$,
and falls off sharply toward higher values. This is indicative of a
parent population that has a distribution of bulk Lorentz factors that
is not peaked at high values. The fastest jet in our sample
(0805$-$077) contains a component moving at $50.6 \pm 2.1 \; c$, which
is likely representative of the upper end of the AGN jet Lorentz
factor distribution.

(viii) The maximum jet speed increases with redshift out to $z \simeq
1$, which reflects a selection bias towards higher Lorentz factor
jets at higher redshifts. At closer distances, the jets do not require
as high beaming factors to meet the flux density selection criteria,
and the probability finding a high luminosity, high Lorentz factor jet is
much lower because of the smaller co-moving volume available.

(ix) We have re-examined the upper envelope in the \ba vs. apparent
luminosity plane first reported by \cite{CLH07} using our new data
set. The envelope is not the result of any observational measurement
bias, and is very similar in shape to an ``aspect curve'' on the plane
traced by a single high-Lorentz factor jet as it is rotated through
different viewing angles.  The statistical probability of having this
envelope, however, is low, in terms of the standard beaming model,
which we have confirmed using Monte Carlo simulations with multiple
random number seeds. It is possible instead that the envelope arises
because of an intrinsic relation between jet speed and luminosity in
the parent population, although deeper samples will likely be needed
to fully investigate this scenario.

The authors acknowledge the contributions of additional members of the
MOJAVE team: Hugh Aller, Margo Aller, Ivan Agudo, Andrei Lobanov,
Alexander Pushkarev, Kirill Sokolovsky, and Rene Vermeulen. Several
students also contributed to this work: Christian Fromm at MPIfR, and
Amy Lankey, Kevin O'Brien, Ben Mohlie, and Nick Mellott at Purdue
University. MLL has been supported under NSF grants AST-0406923 \&
AST-0807860, NASA-\textit{Fermi} grant NNX08AV67G and a grant from the
Purdue Research Foundation. DCH is supported by NSF grant AST-0707693.
TS has been supported in part by the Academy of Finland grant 120516. 
MK has been supported in part by an appointment to the NASA Postdoctoral
Program at the Goddard Space Flight Center, administered by Oak Ridge
Associated Universities through a contract with NASA. Part of this work
was done by YYK and TS during their Alexander von Humboldt fellowships
at the MPIfR.\facility[NRAO(VLBA)]{The National Radio Astronomy
Observatory is a facility of the National Science Foundation operated
under cooperative agreement by Associated Universities, Inc.} This
research has made use of NASA's Astrophysics Data System, and the
NASA/IPAC Extragalactic Database (NED). The latter is operated by the
Jet Propulsion Laboratory, California Institute of Technology, under
contract with the National Aeronautics and Space Administration.

{\it Facilities:} \facility{NRAO (VLBA)}

\clearpage
\begin{landscape}
\begin{deluxetable*}{lcrrrrrrrrrccrrr} 
\tablecolumns{16} 
\tabletypesize{\scriptsize} 
\tablewidth{0pt}  
\tablecaption{\label{velocitytable}Jet Kinematics}  
\tablehead{\colhead{} & \colhead {} &   \colhead {} & 
\colhead{$\langle S\rangle$}  &\colhead{$\langle R\rangle$} & \colhead{$\langle\vartheta\rangle$} & 
\colhead{$\mu$}  &  \colhead{$\beta_{app}$}& \colhead{$\phi$}&   \colhead{$ |\langle\vartheta\rangle - \phi|$}  & \colhead{$ \dot{\mu}_{\parallel} $} &  \colhead{$ \dot{\mu}_{\perp}$}&  \colhead{} & \colhead{} & \colhead{$\Delta \alpha$} &  \colhead{$\Delta \delta$}  \\  
\colhead{Source} & \colhead {I.D.} &  \colhead {N} & 
\colhead{(mJy)} &\colhead{(mas)} & \colhead{(deg)} &\colhead{($\mu$as y$^{-1})$}&  \colhead{$c$}  & 
\colhead{(deg)}& \colhead{(deg)}   & \colhead{($\mu$as y$^{-2})$} &  \colhead{($\mu$as y$^{-2})$}  & \colhead{$T_{ej}$}& \colhead{$T_{mid}$} & \colhead{(mas)}& \colhead{(mas)}  \\  
\colhead{(1)} & \colhead{(2)} & \colhead{(3)} & \colhead{(4)} &  
\colhead{(5)} & \colhead{(6)} & \colhead{(7)} & \colhead{(8)} & 
 \colhead{(9)}& \colhead{(10)}& \colhead{(11)}& \colhead{(12)}& 
\colhead{(13)} & \colhead{(14)} & \colhead{(15)} & \colhead{(16)}    } 
\startdata 
0003$-$066  & 1 & 14  & 176   & 6.7&$ 278.9$ &
 135$\pm$11\tablenotemark{a} & 2.89$\pm$0.24 & 236.9$\pm$8.3 & 42.0$\pm$8.3\tablenotemark{b} &  $-$75.6$\pm$11 &$-$13.1$\pm$14  & \nodata & 2004.37  &0.07 & 0.12 \\ 
  & 3 & 12  & 163   & 1.3&$ 255.2$ &
 101$\pm$9\tablenotemark{a} & 2.17$\pm$0.19 & 263.9$\pm$11 & 8.8$\pm$11 &  29.4$\pm$11 &79.1$\pm$15  & \nodata & 2005.03  &0.05 & 0.07 \\ 
  & 4 & 11  & 507   & 0.8&$ 193.1$ &
 41$\pm$8 & 0.88$\pm$0.17 & 180.8$\pm$9.2 & 12.3$\pm$9.2 &  0.5$\pm$10 &$-$14.1$\pm$6.6  & 1987.3$\pm$7.2 & 2005.03  &0.03 & 0.04 \\ 
  & 5 & 7  & 234   & 0.5&$ 282.6$ &
 123$\pm$16 & 2.64$\pm$0.34 & 229.6$\pm$8.6 & 53.0$\pm$8.9\tablenotemark{b} &  \nodata &\nodata  & \nodata & 2005.62  &0.02 & 0.06 \\ 
0007+106  & 1 & 6  & 336   & 0.4&$ 290.0$ &
 166$\pm$16 & 0.971$\pm$0.094 & 291.1$\pm$2.8 & 1.1$\pm$3.1 &  \nodata &\nodata  & 2003.67$\pm$0.24 & 2005.86  &0.04 & 0.01 \\ 
0016+731  & 1 & 12  & 144   & 1.1&$ 137.3$ &
 87$\pm$5 & 6.74$\pm$0.39 & 152.1$\pm$2.6 & 14.7$\pm$2.8\tablenotemark{b} &  4.5$\pm$2.9 &4.9$\pm$2.3  & \nodata & 2000.63  &0.04 & 0.07 \\ 
0048$-$097  & 1 & 9  & 22   & 1.4&$ 357.9$ &
 344$\pm$71 & \n & 7.8$\pm$4.0 & 10.0$\pm$4.2 &  \nodata &\nodata  & 2003.31$\pm$0.83 & 2004.53  &0.08 & 0.27 \\ 
0059+581  & 2 & 13  & 21   & 1.5&$ 240.8$ &
 197$\pm$14\tablenotemark{a} & 7.28$\pm$0.52 & 253.2$\pm$4.1 & 12.4$\pm$4.3 &  48.7$\pm$13 &$-$1.6$\pm$14  & \nodata & 2004.00  &0.11 & 0.12 \\ 
  & 3 & 10  & 84   & 0.8&$ 234.2$ &
 283$\pm$20 & 10.46$\pm$0.74 & 248.7$\pm$3.8 & 14.5$\pm$4.0\tablenotemark{b} &  $-$19.6$\pm$45 &$-$13.8$\pm$43  & \nodata & 2005.95  &0.07 & 0.06 \\ 
  & 4 & 10  & 155   & 0.6&$ 196.2$ &
 162$\pm$13 & 5.99$\pm$0.48 & 230.7$\pm$4.8 & 34.4$\pm$5.0\tablenotemark{b} &  $-$35.2$\pm$28 &9.9$\pm$30  & \nodata & 2005.95  &0.04 & 0.05 \\ 
  & 5 & 6  & 290   & 0.4&$ 164.7$ &
 300$\pm$23 & 11.09$\pm$0.85 & 166.4$\pm$4.2 & 1.7$\pm$4.4 &  \nodata &\nodata  & 2005.88$\pm$0.17 & 2007.01  &0.02 & 0.02 \\ 
0106+013  & 1 & 6  & 40   & 3.0&$ 237.0$ &
 311$\pm$46 & 26.5$\pm$3.9 & 228.9$\pm$8.4 & 8.1$\pm$8.5 &  \nodata &\nodata  & 1996.2$\pm$2.0 & 2005.33  &0.15 & 0.15 \\ 
  & 2 & 6  & 58   & 1.7&$ 251.7$ &
 286$\pm$45 & 24.4$\pm$3.8 & 263.1$\pm$2.0 & 11.4$\pm$2.2\tablenotemark{b} &  \nodata &\nodata  & \nodata & 2005.33  &0.15 & 0.03 \\ 
  & 3 & 7  & 562   & 0.8&$ 232.6$ &
 249$\pm$9 & 21.22$\pm$0.77 & 239.4$\pm$3.0 & 6.8$\pm$3.1 &  \nodata &\nodata  & 2002.30$\pm$0.14 & 2005.33  &0.02 & 0.05 \\ 
0119+115  & 1 & 9  & 72   & 3.1&$ 6.7$ &
 513$\pm$20 & 17.10$\pm$0.67 & 1.4$\pm$2.4 & 5.2$\pm$2.5 &  \nodata &\nodata  & 1998.43$\pm$0.48 & 2005.03  &0.10 & 0.09 \\ 
\enddata 
\tablenotetext{a}{Component shows significant acceleration.}
\tablenotetext{b}{Component shows significant non-radial motion.}
\tablenotetext{c}{Component shows significant inward motion.}
\tablenotetext{d}{Component is unusually slow (low pattern speed) (see Section \ref{speeddispersion}).}
\tablecomments{Columns are as follows: (1) IAU name (B1950.0); (2) component number; (3) number of fitted epochs; (4) mean flux density;  (5) mean distance from core component; (6) mean position angle with respect to the core component; (7) magnitude of fitted angular velocity vector; (8) fitted speed in units of the speed of light; (9) position angle of velocity vector; (10) offset between mean position angle and velocity vector position angle; (11) acceleration parallel  to velocity direction; (12) acceleration perpendicular to velocity direction; (13) fitted ejection date; (14) date of reference (middle) epoch used for fit; (15) right ascension error of individual epoch positions (mas); (16) declination error of individual epoch positions (mas).}
\end{deluxetable*}
\clearpage
\end{landscape}

\begin{thebibliography}{}

\bibitem[Abdo et al.(2009)]{Fermi3C84} Abdo, A.~A., et al.\ 2009, \apj, 699, 31 

\bibitem[Acciari et al.(2009)]{Acciari09} Acciari, V. A.,  et al. 2009, Science, 325, 447

\bibitem[Alberdi et al.(2000)]{AGM00} Alberdi, A., G{\'o}mez, J.~L., Marcaide, J.~M., Marscher, A.~P., \& P{\'e}rez-Torres, M.~A.\ 2000, \aap, 361, 529 

\bibitem[Alberdi et al.(1997)]{AKG97} Alberdi, A., et al.\ 1997, \aap, 327, 513 

\bibitem[Bach et al.(2005)]{B05} Bach, U., Kadler, M., 
Krichbaum, T.~P., Middelberg, E., Alef, W., Witzel, A., 
\& Zensus, J.~A.\ 2005, Future Directions in High Resolution Astronomy, 340, 30 

\bibitem[Becker et al.(2008)]{BDL08} Becker, P.~A., Das, S., 
\& Le, T.\ 2008, \apjl, 677, L93 

\bibitem[Britzen et al.(2005)]{Britzen05} Britzen, S., et al.\ 2005, \aap, 444, 443 

\bibitem[Britzen et al.(2007)]{B07} Britzen, S., et al.\ 2007, \aap, 472, 763 

\bibitem[Britzen et al.(2008)]{B08} Britzen, S., et al.\ 2008, \aap, 484, 119 

\bibitem[Cara \& Lister(2008)]{CL08} Cara, M., \& Lister, M.~L.\ 2008, \apj, 686, 148 

\bibitem[Ciprini(2009a)]{C09a} Ciprini, S.\ 2009a, The Astronomer's Telegram, 2048, 1 

\bibitem[Ciprini(2009b)]{C09b} Ciprini, S.\ 2009b, The Astronomer's Telegram, 2136, 1

\bibitem[Cohen et al.(2007)]{CLH07} Cohen, M.~H., Lister, 
M.~L., Homan, D.~C., Kadler, M., Kellermann, K.~I., Kovalev, Y.~Y., 
\& Vermeulen, R.~C.\ 2007, \apj, 658, 232 

\bibitem[Cohen et al.(1975)]{C75} Cohen, M.~H., et al.\ 1975, \apj, 201, 249 

\bibitem[Cohen et al.(1977)]{C77} Cohen, M.~H., et al.\ 1977, \nat, 268, 405 

\bibitem[Cooper et al.(2007)]{CLK07} Cooper, N.~J., Lister, 
M.~L., \& Kochanczyk, M.~D.\ 2007, \apjs, 171, 376 

\bibitem[D'Arcangelo et al.(2007)]{Darc07} D'Arcangelo, F.~D., 
et al.\ 2007, \apjl, 659, L107 

\bibitem[Dhawan et al.(1998)]{DKR98} Dhawan, V., Kellermann, 
K.~I., \& Romney, J.~D.\ 1998, \apjl, 498, L111 

\bibitem[Donoso et al.(2009)]{DBK09} Donoso, E., Best, P.~N., 
\& Kauffmann, G.\ 2009, \mnras, 392, 617 

\bibitem[Drinkwater et al.(1997)]{DWF97} Drinkwater, M.~J., et al.\ 1997, \mnras, 284, 85 

\bibitem[Falomo(1996)]{Falomo96} Falomo, R.\ 1996, \mnras, 283, 
241 

\bibitem[Feissel-Vernier(2003)]{FV03} Feissel-Vernier, M.\ 2003, \aap, 403, 105 

\bibitem[G\'omez et al.(1997)]{GMM97} G\'omez, J.~L., Mart\'{\i}, 
J.~M.~A., Marscher, A.~P., Ib\'a\~nez, J.~M.~A.,
\& Alberdi, A.\ 1997, \apjl, 482, L33 

\bibitem[G\'omez et al.(1995)]{GMM95} G\'omez, J.~L., Mart\'{\i}, 
J.~M.~A., Marscher, A.~P., Ib\'a\~nez, J.~M.~A., 
\& Marcaide, J.~M.\ 1995, \apjl, 449, L19 

\bibitem[Gong(2008)]{G08} Gong, B.\ 2008, \mnras, 389, 315

\bibitem[Healey et al.(2008)]{Healey08} Healey, S.~E., et al.\ 
2008, \apjs, 175, 97 

\bibitem[Helmboldt et al.(2007)]{Helm07} Helmboldt, J.~F., et 
al.\ 2007, \apj, 658, 203 

\bibitem[Homan et al.(2009)]{Homan09}
Homan, D.~C., et al. 2009, \apj, in press; arXiv:0909.5102 (Paper~VII)

\bibitem[Homan \& Lister(2006)]{HL06} Homan, D.~C., \& Lister, M.~L.\ 2006, \aj, 131, 1262 

\bibitem[Homan et al.(2003)]{H03}
Homan, D.~C., et al.\ 2003, \apjl, 589, L9 

\bibitem[Homan et al.(2001)]{HOW01} Homan, D.~C., Ojha, R., 
Wardle, J.~F.~C., Roberts, D.~H., Aller, M.~F., Aller, H.~D., 
\& Hughes, P.~A.\ 2001, \apj, 549, 840 

\bibitem[Homan et al.(2002)]{HOW02} Homan, D.~C., Ojha, R., 
Wardle, J.~F.~C., Roberts, D.~H., Aller, M.~F., Aller, H.~D., \& Hughes, 
P.~A.\ 2002, \apj, 568, 99 

\bibitem[Homan et al.(2006)]{H06} Homan, D.~C., et al.\ 
2006, \apjl, 642, L115 

\bibitem[Istomin \& Pariev(1996)]{IP96}
Istomin, Y.~N., \& Pariev, V.~I.\ 1996, \mnras, 281, 1 

\bibitem[Jammalamadaka \& Sengupta(2001)]{JS01}
Jammalamadaka, S. R., \& Sengupta, A., 2001, Topics in Circular
Statistics, World Scientific Publishing, 15

\bibitem[Jorstad et al.(2004)]{JML04}
Jorstad, S.~G., Marscher, A.~P., Lister, M.~L., Stirling, A.~M.,
Cawthorne, T.~V., G{\'o}mez, J.-L., \& Gear, W.~K.\ 2004, \aj, 127,
3115 

\bibitem[Jorstad et al.(2001)]{J01}
Jorstad, S.~G., Marscher, A.~P., Mattox, J.~R., Aller, M.~F., Aller,
H.~D., Wehrle, A.~E., \& Bloom, S.~D.\ 2001, \apj, 556, 738 

\bibitem[Jorstad et al.(2005)]{J05}
Jorstad, S.~G., et al.\ 2005, \aj, 130, 1418 

\bibitem[Kadler et al.(2004)]{kad04b}
Kadler, M., Kerp, J., Ros, E., Falcke, H., Pogge, R.~W., \& Zensus, J.~A.\ 2004, \aap, 420, 467

\bibitem[Kadler et al.(2007)]{KMO07}
Kadler, M., Ojha, R., Tingay, S., Lovell, J., \& TANAMI collaboration 2007, American Astronomical Society Meeting Abstracts, 211, \#04.13 

\bibitem[Kadler et al.(2004)]{k04b}
Kadler, M., Ros, E., Lobanov, A.~P., Falcke, H., \& Zensus, J.~A.\ 2004, \aap, 426, 481

\bibitem[Kadler et al.(2008)]{K08}
Kadler, M., et al.\ 2008, \apj, 680, 867 

\bibitem[Kellermann et al.(1998)]{K98}
Kellermann, K.~I., Vermeulen, R.~C., Zensus, J.~A., \& Cohen, M.~H.\
1998, \aj, 115, 1295 

\bibitem[Kellermann et al.(2004)]{KL04}
Kellermann, K.~I.  et al., 2004, \apj, 609, 539 

\bibitem[Kovalev et al.(1999)]{KNK99}
Kovalev, Y.~Y., Nizhelsky, N.~A., Kovalev, Y.~A., Berlin, A.~B.,
Zhekanis, G.~V., Mingaliev, M.~G., \& Bogdantsov, A.~V.\ 1999, \aaps,
139, 545 

\bibitem[Kovalev et al.(2005)]{KKL05}
Kovalev, Y.~Y., et al.\ 2005, \aj, 130, 2473 

\bibitem[Kovalev et al.(2007)]{KLH07}
Kovalev, Y.~Y., Lister, M.~L., Homan, D.~C., \& Kellermann, K.~I.\ 2007,
\apjl, 668, L27 

\bibitem[Kovalev et al.(2008)]{KLP08}
Kovalev, Y.~Y., Lobanov, A.~P., Pushkarev, A.~B., \& Zensus, J.~A.\
2008, \aap, 483, 759

\bibitem[Kovalev et al.(2009)]{KAA09}
Kovalev, Y.~Y., et al.\ 2009, \apjl, 696, L17 

\bibitem[Lawrence et al.(1996)]{LZR96}
Lawrence, C.~R., Zucker, J.~R., Readhead, A.~C.~S., Unwin, S.~C.,
Pearson, T.~J., \& Xu, W.\ 1996, \apjs, 107, 541 

\bibitem[Lister(2001)]{L01a} Lister, M.~L.\ 2001, \apj, 562, 
208 

\bibitem[Lister(2003)]{Lister03} Lister, M.~L.\ 2003, \apj, 599, 
105 

\bibitem[Lister \& Homan(2005)]{LH05} Lister, M.~L., \& Homan, D.~C.\ 2005, \aj, 130, 1389 

\bibitem[Lister et al.(2009)]{LHK09} Lister, M.~L., Homan, D.~C., Kadler, M., Kellermann, K.~I., Kovalev, Y.~Y.,
Ros, E., Savolainen, T., \& Zensus, J.~A.\ 2009, \apjl, 696, L22 

\bibitem[Lister et al.(2003)]{LKV03} Lister, M.~L., 
Kellermann, K.~I., Vermeulen, R.~C., Cohen, M.~H., Zensus, J.~A., 
\& Ros, E.\ 2003, \apj, 584, 135 

\bibitem[Lister \& Marscher(1997)]{LM97} Lister, M.\ L.\ \&  Marscher,
A.\ P.\ 1997, \apj, 476, 572

\bibitem[Lister et al.(2001a)]{LTM01} Lister, M.~L., Tingay, 
S.~J., Murphy, D.~W., Piner, B.~G., Jones, D.~L., 
\& Preston, R.~A.\ 2001a, \apj, 554, 948 

\bibitem[Lister et al.(2001b)]{LTP01} Lister, M.~L., Tingay, 
S.~J., \& Preston, R.~A.\ 2001b, \apj, 554, 964 

\bibitem[Lister \& Smith(2000)]{LS00} Lister, M.~L., \& Smith, P.~S.\ 2000, \apj, 541, 66 

\bibitem[Lister et al.(2009)]{LAA09} Lister, M. ~L., et al., 2009,
\aj, 137, 3718 (Paper~V)

\bibitem[Lobanov(1998)]{Lobanov98} Lobanov, A.~P.\ 1998, \aap, 330, 79 

\bibitem[Ly et al.(2007)]{LWJ07}
Ly, C., Walker, R.~C., \& Junor, W.\ 2007, \apj, 660, 200 

\bibitem[Mandal \& Chakrabarti(2008)]{MC08} Mandal, S., \& Chakrabarti, S.~K.\ 2008, \apjl, 689, L17 

\bibitem[Marscher et al.(2002)]{MJM02} Marscher, A.~P., 
Jorstad, S.~G., Mattox, J.~R., \& Wehrle, A.~E.\ 2002, \apj, 577, 85 

\bibitem[Marscher et al.(1991)]{MZS91} Marscher, A.~P., 
Zhang, Y.~F., Shaffer, D.~B., Aller, H.~D., 
\& Aller, M.~F.\ 1991, \apj, 371, 491 

\bibitem[Marscher et al.(2008)]{MJ08} Marscher, A.~P., et 
al.\ 2008, \nat, 452, 966 

\bibitem[McIntosh et al.(1999)]{MRR99}
McIntosh, D.~H., Rieke, M.~J., Rix, H.-W., Foltz, C.~B., \& Weymann, R.~J.\ 1999, \apj, 514, 40 

\bibitem[Moellenbrock (1998)]{Moe98} Moellenbrock, G. 1998,
Ph.D. Thesis, Brandeis University


\bibitem[Perley et al.(1982)]{PFJ82} Perley, R.~A., Fomalont, 
E.~B., \& Johnston, K.~J.\ 1982, \apjl, 255, L93 

\bibitem[Perlman et al.(2002)]{PSC02} Perlman, E.~S., Stocke, 
J.~T., Carilli, C.~L., Sugiho, M., Tashiro, M., Madejski, G., Wang, Q.~D., 
\& Conway, J.\ 2002, \aj, 124, 2401 

\bibitem[Perucho 
\& Mart{\'{\i}}(2007)]{PM07} Perucho, M., \& Mart{\'{\i}}, J.~M.\ 2007, \mnras, 382, 526 

\bibitem[Piner et al.(2007)]{PMF07} Piner, B.~G., Mahmud, M., 
Fey, A.~L., \& Gospodinova, K.\ 2007, \aj, 133, 2357 

\bibitem[Schilizzi et al.(1975)]{SCM75} Schilizzi, R.~T., 
Cohen, M.~H., Shaffer, D.~B., Kellermann, K.~L., Swenson, G.~W., Jr., Yen, 
J.~L., Rinehart, R., \& Romney, J.~D.\ 1975, \apj, 201, 263 

\bibitem[Shaffer et al.(1975)]{S75} Shaffer, D.~B., Cohen, 
M.~H., Schilizzi, R.~T., Kellermann, K.~I., Swenson, G.~W., Jr., Yen, 
J.~L., Rinehart, R., \& Romney, J.~D.\ 1975, \apj, 201, 256 

\bibitem[Shepherd(1997)]{1997ASPC..125...77S} Shepherd, M.~C.\ 1997, 
Astronomical Data Analysis Software and Systems VI, 125, 77 

\bibitem[Small et al.(1997)]{SSS97} Small, T.~A., Sargent, W.~L.~W., \& Steidel, C.~C.\ 1997, \aj, 114, 2254 

\bibitem[Steidel \& Sargent(1991)]{SS91} Steidel, C.~C., \&
Sargent, W.~L.~W.\ 1991, \apj, 382, 433 	

\bibitem[Stirling et al.(2003)]{SCS03} Stirling, A. M., et al., 2003, \mnras, 341, 405

\bibitem[Tadhunter et al.(1993)]{TMS93} Tadhunter, C.~N., 
Morganti, R., di Serego-Alighieri, S., Fosbury, R.~A.~E., 
\& Danziger, I.~J.\ 1993, \mnras, 263, 999 

\bibitem[Taylor et al.(1996)]{TVR96} Taylor, G.~B., 
Vermeulen, R.~C., Readhead, A.~C.~S., Pearson, T.~J., Henstock, D.~R., 
\& Wilkinson, P.~N.\ 1996, \apjs, 107, 37 

\bibitem[Torniainen et al.(2005)]{TTT05} Torniainen, I., Tornikoski, M., Ter{\"a}sranta, H., Aller, M.~F., \& Aller, H.~D.\ 2005, \aap, 435, 839 

\bibitem[Ulvestad et al.(1981)]{UJP81} Ulvestad, J., Johnston, K., Perley, R., \& Fomalont, E.\ 1981, \aj, 86, 1010 
 
\bibitem[Vermeulen \& Cohen(1994)]{VC94} Vermeulen, R.~C., \& Cohen, M.~H.\ 1994, \apj, 430, 467 

\bibitem[Vermeulen et al.(2003)]{VRK03} Vermeulen, R.~C., Ros, E., Kellermann, K.~I., Cohen, M.~H., Zensus, J.~A., \& van Langevelde, H.~J.\ 2003, \aap, 401, 113 

 \bibitem[V{\'e}ron-Cetty \& V{\'e}ron(2006)]{VCV06} V{\'e}ron-Cetty,
 M.-P., \& V{\'e}ron, P.\ 2006, \aap, 455, 773

\bibitem[Wagner et al.(2009)]{Wagner09} Wagner, R.~M., et al.\ 
2009, arXiv:0907.1465 

\bibitem[Walker et al.(2000)]{WDR00} Walker, R.~C., Dhawan, V., Romney, J.~D., Kellermann, K.~I., \& Vermeulen, R.~C.\ 2000, \apj, 530, 233

\bibitem[Wardle \& Aaron(1997)]{WA97} Wardle, J.~F.~C., \& Aaron, S.~E.\ 1997, \mnras, 286, 425 

\bibitem[White et al.(1988)]{WJW88} White, G.~L., Jauncey, 
D.~L., Wright, A.~E., Batty, M.~J., Savage, A., Peterson, B.~A., 
\& Gulkis, S.\ 1988, \apj, 327, 561 

\bibitem[Wiik et al.(2001)]{WVL01} Wiik, K., Valtaoja, E., \& Lepp{\"a}nen, K.\ 2001, \aap, 380, 72 

\bibitem[Wrobel(1984)]{Wro84} Wrobel, J.~M.\ 1984, \apj, 284, 531 

\bibitem[Zensus et al.(2002)]{Z02} Zensus, A., Ros, E., Kellermann, K.~I., Cohen, M.~H., Vermeulen, R.~C.\ 2003, \aj, 124, 662

\end{thebibliography}
\end{document}